\documentclass[manuscript]{acmart}
\usepackage{graphicx}
\usepackage{lineno}

\usepackage{adjustbox}
\usepackage{listings}
\usepackage{xcolor}
\usepackage{algorithm2e}
\usepackage{pifont}

\definecolor{lightgray}{rgb}{9, 9, .9}
\definecolor{darkgray}{rgb}{.4, .4, .4}
\definecolor{purple}{rgb}{0.65, 0.12, 0.82}
\definecolor{gray}{rgb}{0.4,0.4,0.4}
\definecolor{line-numbers}{rgb}{0.4,0.4,0.4}
\definecolor{tags}{rgb}{1, 0, 0}
\definecolor{darkblue}{rgb}{0.0,0.0,0.6}
\definecolor{cyan}{rgb}{0.0,0.6,0.6}
\definecolor{highlight}{HTML}{ffffff}

\usepackage{soul}

\newcommand{\hlc}[2][yellow]{{%
    \colorlet{foo}{#1}%
    \sethlcolor{foo}\hl{#2}}%
}

\lstdefinelanguage{Groovy}
{
  sensitive=true,%
    morecomment=[l]//,% 
  morecomment=[s]{/}{/},% 
  morestring=[b]",% 
  morestring=[b]',% 
  stringstyle=\color{black},
  identifierstyle=\color{darkblue},
  keywordstyle=\color{cyan},
  morekeywords={abstract,any,as,boolean,break,byte,case,catch,char,
  class, const,continue,def,default,do,double,else,extends,false,final,finally, float,for,goto,if,implements,import,instanceof,in,int,interface,label, long,native,new,null,package,private,protected,public,return,short, static,strictfp,super,switch,synchronized,this,throw,throws,transient, true,try,void,volatile,while,with}% list your attributes here
}
\usepackage[T1]{fontenc}
\usepackage{listings,beramono}
\usepackage{tikz}
\makeatletter
\newenvironment{btHighlight}[1][]
{\begingroup\tikzset{bt@Highlight@par/.style={#1}}\begin{lrbox}{\@tempboxa}}
{\end{lrbox}\bt@HL@box[bt@Highlight@par]{\@tempboxa}\endgroup}

\newcommand\btHL[1][]{%
  \begin{btHighlight}[#1]\bgroup\aftergroup\bt@HL@endenv%
}
\def\bt@HL@endenv{%
  \end{btHighlight}%   
  \egroup
}
\newcommand{\bt@HL@box}[2][]{%
  \tikz[#1]{%
    \pgfpathrectangle{\pgfpoint{1pt}{0pt}}{\pgfpoint{\wd #2}{\ht #2}}%
    \pgfusepath{use as bounding box}%
    \node[anchor=base west, fill=orange!30,outer sep=0pt,inner xsep=1pt, inner ysep=0pt, rounded corners=3pt, minimum height=\ht\strutbox+1pt,#1]{\raisebox{1pt}{\strut}\strut\usebox{#2}};
  }%
}
\makeatother

\lstdefinestyle{Groovy}{
    language={Groovy}, %basicstyle=\scriptsize\ttfamily,
    moredelim=**[is][\btHL]{`}{`},
    moredelim=**[is][{\btHL[fill=green!30]}]{*}{*},
    moredelim=**[is][{\btHL[fill=cyan!30]}]{~}{~},
    extendedchars=true,
}

\usepackage{qtree}
\usepackage{multicol}        
\usepackage{makecell}
\usepackage{url}   %%%%%%%%%%%%%%%%%%%%%%%%%%%%%%%%%%%%%%%%%%%%%%%%%%%%%%%%%%%%%%%%%%%%%
\makeatletter

\newcommand\PrologPredicateStyle{}
\newcommand\PrologVarStyle{}
\newcommand\PrologAnonymVarStyle{}
\newcommand\PrologAtomStyle{}
\newcommand\PrologOtherStyle{}
\newcommand\PrologCommentStyle{}

\newif\ifpredicate@prolog@
\newif\ifwithinparens@prolog@

\lst@SaveOutputDef{`_}\underscore@prolog

\newcount\currentchar@prolog

\newcommand\@testChar@prolog%
{%
  \ifnum\lst@mode=\lst@Pmode%
    \detectTypeAndHighlight@prolog%
  \else
    \ifwithinparens@prolog@%
      \detectTypeAndHighlight@prolog%
    \fi
  \fi
  \global\predicate@prolog@false%
}

\newcommand\detectTypeAndHighlight@prolog
{%
  \def\lst@thestyle{\PrologAtomStyle}%
  \ifpredicate@prolog@%
    \def\lst@thestyle{\PrologPredicateStyle}%
  \else
    \expandafter\splitfirstchar@prolog\expandafter{\the\lst@token}%
    \expandafter\ifx\@testChar@prolog\underscore@prolog%
      \ifnum\lst@length=1%
        \let\lst@thestyle\PrologAnonymVarStyle%
      \else
        \let\lst@thestyle\PrologVarStyle%
      \fi
    \else
      \currentchar@prolog=65
      \loop
        \expandafter\ifnum\expandafter`\@testChar@prolog=\currentchar@prolog%
          \let\lst@thestyle\PrologVarStyle%
          \let\iterate\relax
        \fi
        \advance \currentchar@prolog by 1
        \unless\ifnum\currentchar@prolog>90
      \repeat
    \fi
  \fi
}
\newcommand\splitfirstchar@prolog{}
\def\splitfirstchar@prolog#1{\@splitfirstchar@prolog#1\relax}
\newcommand\@splitfirstchar@prolog{}
\def\@splitfirstchar@prolog#1#2\relax{\def\@testChar@prolog{#1}}

\def\beginlstdelim#1#2%
{%
  \def\endlstdelim{\PrologOtherStyle #2\egroup}%
  {\PrologOtherStyle #1}%
  \global\predicate@prolog@false%
  \withinparens@prolog@true%
  \bgroup\aftergroup\endlstdelim%
}

\newcommand\lang@prolog{Prolog-pretty}
\expandafter\lst@NormedDef\expandafter\normlang@prolog%
  \expandafter{\lang@prolog}

\expandafter\expandafter\expandafter\lstdefinelanguage\expandafter%
{\lang@prolog}
{%
  language            = Prolog,
  keywords            = {},      % reset all preset keywords
  showstringspaces    = false,
  alsoletter          = (,
  alsoother           = @\$,
  moredelim           = **[is][\beginlstdelim{(}{)}{(}{)}],
  MoreSelectCharTable =
    \lst@DefSaveDef{`(}\opparen@prolog{\global\predicate@prolog@true\opparen@prolog},
}

\newcommand\@ddedToOutput@prolog\relax
\lst@AddToHook{Output}{\@ddedToOutput@prolog}

\lst@AddToHook{PreInit}
{%
  \ifx\lst@language\normlang@prolog%
    \let\@ddedToOutput@prolog\@testChar@prolog%
  \fi
}

\lst@AddToHook{DeInit}{\renewcommand\@ddedToOutput@prolog{}}

\makeatother
\definecolor{PrologPredicate}{RGB}{000,031,255}
\definecolor{PrologVar}      {RGB}{024,021,125}
\definecolor{PrologAnonymVar}{RGB}{000,127,000}
\definecolor{PrologAtom}     {RGB}{186,032,032}
\definecolor{PrologComment}  {RGB}{063,128,127}
\definecolor{PrologOther}    {RGB}{000,000,000}

\renewcommand\PrologPredicateStyle{\color{PrologPredicate}}
\renewcommand\PrologVarStyle{\color{PrologVar}}
\renewcommand\PrologAnonymVarStyle{\color{PrologAnonymVar}}
\renewcommand\PrologAtomStyle{\color{PrologAtom}}
\renewcommand\PrologCommentStyle{\itshape\color{PrologComment}}
\renewcommand\PrologOtherStyle{\color{PrologOther}}

\lstdefinestyle{Prolog-pygsty}
{
  language     = Prolog-pretty,
  upquote      = true,
  moredelim=**[is][\btHL]{`}{`},
    moredelim=**[is][{\btHL[fill=green!30]}]{*}{*},
    moredelim=**[is][{\btHL[fill=cyan!30]}]{~}{~},
  stringstyle  = \PrologAtomStyle,
  commentstyle = \PrologCommentStyle,
  literate     =
    {:-}{{\PrologOtherStyle :-}}2
    {,}{{\PrologOtherStyle ,}}1
    {.}{{\PrologOtherStyle .}}1
}

\usepackage{textcomp}
\usepackage{pgfplots}
\pgfplotsset{width=8cm,compat=1.9}

\usepackage{fncychap}

\usepackage[toc,page]{appendix}

\usepackage{multirow}

\begin{document}

\title{A Model-Driven-Engineering Approach for Detecting Privilege Escalation in IoT Systems}
\begin{comment}
\author{Ben Trovato}
\authornote{Both authors contributed equally to this research.}
\email{trovato@corporation.com}
\orcid{1234-5678-9012}
\author{G.K.M. Tobin}
\authornotemark[1]
\email{webmaster@marysville-ohio.com}
\affiliation{%
  \institution{Institute for Clarity in Documentation}
  \streetaddress{P.O. Box 1212}
  \city{Dublin}
  \state{Ohio}
  \country{USA}
  \postcode{43017-6221}
}
\end{comment}

\author{Atheer Abu Zaid}
\affiliation{%
  \institution{Ryerson University}
  %\streetaddress{1 Th{\o}rv{\"a}ld Circle}
  \city{Toronto}
  \country{Canada}}
\email{aabuzaid@ryerson.ca}

\author{Manar H. Alalfi}
\affiliation{%
  \institution{Ryerson University}
  \city{Toronto}
  \country{Canada}}
\email{manar.alalfi@ryerson.ca}

\author{Ali Miri}
\affiliation{%
  \institution{Ryerson University}
  \city{Toronto}
  \country{Canada}}
\email{Ali.Miri@ryerson.ca}

\begin{comment}
\author{Aparna Patel}
\affiliation{%
 \institution{Rajiv Gandhi University}
 \streetaddress{Rono-Hills}
 \city{Doimukh}
 \state{Arunachal Pradesh}
 \country{India}}

\author{Huifen Chan}
\affiliation{%
  \institution{Tsinghua University}
  \streetaddress{30 Shuangqing Rd}
  \city{Haidian Qu}
  \state{Beijing Shi}
  \country{China}}

\author{Charles Palmer}
\affiliation{%
  \institution{Palmer Research Laboratories}
  \streetaddress{8600 Datapoint Drive}
  \city{San Antonio}
  \state{Texas}
  \country{USA}
  \postcode{78229}}
\email{cpalmer@prl.com}

\author{John Smith}
\affiliation{%
  \institution{The Th{\o}rv{\"a}ld Group}
  \streetaddress{1 Th{\o}rv{\"a}ld Circle}
  \city{Hekla}
  \country{Iceland}}
\email{jsmith@affiliation.org}

\author{Julius P. Kumquat}
\affiliation{%
  \institution{The Kumquat Consortium}
  \city{New York}
  \country{USA}}
\email{jpkumquat@consortium.net}
\end{comment}

\renewcommand{\shortauthors}{Atheer Abu Zaid, Manar H. Alalfi, Ali Miri}

\begin{abstract}
\hlc[highlight]{Software vulnerabilities in access control models can represent a serious threat in a system. 
In fact, OWASP lists broken access control as number 5 in severity among the top 10 vulnerabilities. 
In this paper, we study the permission model of an emerging Smart-Home platform, SmartThings, and explore an approach that detects privilege escalation in its permission model.}
%The first approach applies static analysis to extract vulnerabilities by pattern matching.
Our approach is based on Model Driven Engineering (MDE) in addition to static analysis.
%The second approach complements the static analysis-based approach which cannot analyze the semantic itself.
%MDE-based
This approach allows for better coverage of privilege escalation detection than static analysis alone, and takes advantage of analyzing free-form text that carries extra permissions details.
Our experimental results demonstrate a very high accuracy for detecting over-privilege vulnerabilities in IoT applications.
\end{abstract}



%\acmConference{Some conference}
\maketitle

%\linenumbers


\section{Introduction}
IoT - term coined in 1999 by Kevin Ashton - emerged to describe when the Internet is used to connect objects as end users, rather than just people \cite{gubbi2013internet}. 
IoT has a variety of definitions by different groups in the academia and the industry \cite{gubbi2013internet,madakam2015internet,wortmann2015internet}. 
A common understanding of IoT is the interconnection of uniquely identifiable, ubiquitous, sensing/actuating capable, programmable and self configurable things over the internet. These things share data and services without the need for human involvement \cite{ieee2015towards}.
%\cite{madakam2015internet}. 
%These objects include embedded devices, gadgets, appliances, cars and anything that can be connected and identified on the Internet. 
IoT is now widely adopted in many important domains, such as in healthcare systems, smart cities, smart homes and autonomous cars \cite{soumyalatha2016study} \cite{8325597}.
Employing IoT systems in cities has proved to be useful, such as in improving transportation by gathering data and analyzing it \cite{soumyalatha2016study}. 
It is also important in smart homes to automate everyday tasks.
Despite popularity of IoT systems in practice, their design and implementation maturity is still in early stages.  
In particular, IoT systems suffer from many security vulnerabilities, some of which   have already been addressed in other types of systems.
The issues of interest to us in this paper are those related to access control, which is an important component of IoT systems. 
Access control models provide rules that administer and constrain how objects access and interact with each other. 
We will discuss access control models in more details later.

%\section{Smart-Homes}
An important application of IoT is smart homes. A smart home is a system in a residence that allows the household to monitor and control home devices and appliances \cite{alam2012review,alhanahnah2019advanced}. 
In \cite{zhou2019discovering}, \emph{Zhou et al.} provide a concise description of the current trend in smart home platforms. They demonstrate that cloud-based smart home platforms typically have three components: the \emph{cloud backend}, the \emph{physical IoT devices} and the \emph{mobile app}. 
The cloud is responsible for identity management, executing applications and home automation. Apps are executed in the cloud to ensure that the remote commands are sent by authenticated users. 
IoT devices have sensors for monitoring and actuators for executing commands. The devices can either be cloud-connected devices or hub-connected devices. 
The last component, the mobile app, provides the user with an interface that connects IoT devices to the smart home to specify the desired home automation.

Some of the most popular smart home platforms are: Samsung's SmartThings, Apple HomeKit, Vera Control's Vera3, Google's Weave/Brillo, AllSeen Alliance's AllJoyn, Amazon Alexa, Google Assistant and IFTTT \cite{FernandesJP16,alhanahnah2019advanced}. 

Programmable Smart home platforms have emerged over the last years and have provided users with broad benefits. These platforms provide third-party developers with the opportunity to develop apps for compatible devices. Unfortunately, this flexibility that apps can be developed by third-party developers introduced new risks and attacks into these platforms. In this paper, the platform of interest is Samsung's SmartThings. Similar to other programmable platforms, it is cloud-based and provides a programming framework for third-party developers \cite{FernandesJP16}. Furthermore, SmartThings shares two other common features with other platforms: authorization and authentication through the capability model, and the support of event-driven processing. Thus, analysis done on this platform can be further extended to other similar working platforms.
\section{Motivation}
One of the fundamental software security principles is the least privilege principle \cite{stallings}. 
This principle indicates that an entity should be provided with the least privileges it needs to perform its function. 
The entity should not be able to access a resource if not given permission explicitly. 
Evading this principle might result in misuse of resources and other security misuse implications.

The vulnerability we are interested in this paper is over-privilege in IoT applications, which maps to the "Broken Access Control" in OWASP's top 10 IoT vulnerability list, that is listed as number 5 in importance. OWASP top 10  vulnerabilities is a valuable resource for researchers and businesses to consider when auditing systems for security \cite{owasp}. 

Design flaws in the SmartThings platform have resulted in apps acquiring access to more resources than required \cite{fernandes2017security}. Another issue in the design of the SmartThings permission model is the ability to gain elevated access to unauthorized resources in the devices. This is called privilege escalation. Detecting privilege escalation issues in software before publishing it is of great importance for protecting the users. 
Threats in SmartThings that allow for extra access to resources motivate us to improve the detection of vulnerabilities concerning the access control model.
%We have taken two approaches to detection of privilege escalation.
%The first applies static analysis techniques to do the detection.
%In the second approach, we apply model driven engineering (MDE), a methodology used to abstract the system and extract models that represent it at a high level, through a process called meta-modeling \cite{schmidt2006model}. Those models and constraints allow for the process of model checking. We chose Prolog for the model checking, through applying Prolog rules and queries against Prolog facts to conduct the detection of over-privilege.
%We demonstrate how the MDE approach is capable of combining multiple sources of information for better understanding of permissions granted to the software.

%In a previous work \cite{ours}, we employed static analysis for the detection of over-privilege in SmartThings applications. We designed an approach that looks for vulnerability fingerprints. The search for fingerprints is done by analyzing patterns of vulnerabilities in the software. Patterns are inferred from the SmartThings framework documentation.

In this paper, %we extend our earlier work that looked for vulnerability fingerprints as part of a static analysis for the detection of over-privilege in SmartThings applications. 
we present an MDE approach that provides a comprehensive solution to detect the different over-privilege access scenarios in SmartThings applications. 
%Our analysis specifically targets cases where the description of the app and information present to the user are ambiguous or not sufficient to explain the extent to which the resources will be accessed.
%our proposed approach complements the first one by extending the coverage of privilege escalation to include cases %detection in SmartApps. In cases 2 \& 3, the software is authorized to access the resources, however,
%where the description of the app and information present to the user are ambiguous or not sufficient to explain the extent to which the resources will be accessed.
To address these cases of over-privilege our paper presents the following contributions:
\begin{itemize}
\item We designed and developed an approach and a tool that employs MDE in addition to static analysis to translate multiple sources of information in the app into permission rules. Unlike other existing techniques \cite{zhang2018homonit,TianZLWUGT17,einarsson2017smarthomeml,FernandesJP16}, our approach is comprehensive in detection and coverage of privilege escalation cases in SmartThings.
\item We designed a meta-model that is used to extract permission rules in IoT apps. Once the permission rules are extracted in the form of Prolog facts, Prolog is used to check if the app conforms to the permission meta-model.
\item An experiment that attempts to address the following research questions:
\begin{enumerate}
    \item \hlc[highlight]{RQ1: What is the performance of our approach? How scalable is our analysis when applied to a large dataset of SmartThings apps? Can our automated analysis report the 3 cases of over-privilege in SmartThings apps?}
    \item \hlc[highlight]{RQ2: How does privilege escalation detection performed by ChYP} \cite{chyp} \hlc[highlight]{compare to the MDE approach presented in this paper?}
    \item \hlc[highlight]{RQ3: How effective is our tool in detecting privilege escalation in SmartThings apps when evaluated for precision and recall?}
\end{enumerate}

%\item A tool to detect privilege escalation in SmartThings apps. Unlike other existing techniques \cite{zhang2018homonit,TianZLWUGT17,einarsson2017smarthomeml, FernandesJP16}, our approach is comprehensive in coverage of privilege escalation cases in SmartThings.
\end{itemize}
\section{Background}
%\section{SmartThings: A Smart Home Platform}
The SmartThings platform is a cloud-based system that provides a programming framework for third-party developers. Apps written by the developers are executed in a sand-boxed environment on the cloud to restrict the allowed operations to third-party developers. 
A central device in the smart-home, \emph{the hub}, connects the apps on the cloud with the smart devices. 
\emph{SmartApps} in this framework are applications that can communicate with smart devices. These communications are based on a permission model, also called the capability model.

Listing \ref{lst1} demonstrates the structure of the SmartThings SmartApp. A SmartApp has four general sections: \emph{Definition}, \emph{Preferences}, \emph{Predefined Callbacks} and \emph{Event Handlers} \cite{anatomy}. 
The \emph{Definition} section includes meta-data about the app that appear in the mobile app UI, including the description of the app. 
Capabilities and input requests are provided in the \emph{Preferences} section. 
In listing \ref{lst1}, the app requests two capabilities: motionSensor and lock.
Once the app is installed, the user must provide the app with access to compatible devices for each requested capability. 
Two common \emph{Predefined Callbacks} are \emph{`installed'} and \emph{`updated'}. 
They are called automatically when the app is installed and updated, respectively. 
Event subscription is typically set up in those callbacks. 
It allows the app to listen to events from the device. 

\begin{lstlisting}[style=Groovy, caption={SmartApp Structure}, label={lst1}]
definition
(   name: "",
    namespace: "",
    author: "",
    description: "",
    ...)
preferences
{   section("Select devices") 
    { input "themotion", "capability.motionSensor", title: "Select a motion sensor"
      input "thelock", "capability.lock", title: "Select a lock" }}
def installed() { initialize() }
def updated() { unsubscribe()  initialize() }
def initialize() { subscribe themotion, "motion.active", activeHandler }
def activeHandle(evt) { thelock.unlock() }
\end{lstlisting}

The \emph{Event Handler} \emph{activeHandler} in Listing \ref{lst1} unlocks the device when an \emph{`active'} event is triggered. Lastly, commands and attributes are usually invoked in \emph{Event Handlers}, such as the \emph{`unlock'} command being invoked in \emph{activeHandler}.A capability model represents how smart devices are capable of performing functionalities. The capability structure that this model relies on has two components; \emph{commands} and \emph{attributes}. The over-privilege vulnerability in the SmartThings platform will be discussed further in this chapter. A capability can have multiple commands and attributes to support it. 
%As demonstrated in Figure \ref{fig:thecapability}. 
Attributes hold the state of the device in regard to some property, while commands are used to actuate actions on the device. A device usually has more than one capability, such as having a battery, a light, some sensors or any resource that a smart device might need to perform its job. Apps written to interact with these devices have to specify and request which of the supported capabilities they want to access. As long as a device provides all the needed capabilities of an app, the app is compatible to access this device. This is a restriction of the capability model. 

On installation of the app, the end user is given the option to choose which of the compatible devices they want to be controlled by this app. 
Typically, the set of capabilities that the app requested of the devices are not shown to the user. 
This is done in the code and is not explicitly presented to the user.
This may result in users not being able to detect an app being over-privileged.
%\begin{minipage}{\linewidth}

%\end{minipage}
%\begin{figure}[h!]
%\centerline{\includegraphics[width=.2\textwidth]{Figures/thecapability.png}}
%\caption{Composition of a Capability in SmartThings}
%\label{fig:thecapability}
%\end{figure}
\subsection{Existing Access Control Models in Smart Homes}
A survey by Fernandes \emph{et al.} \cite{fernandes2017security} on access control in IoT devices, and in particular smart home programming frameworks, illustrated that granularity in access control varies from all-or-nothing to very fine grained.  Apple’s HomeKit framework requires access at the home-level granularity (all-or-nothing), meaning iOS apps will always be over-privileged. AllJoyn framework offers access control on a fine-grained level (they call it member level), but recommends it to be at the interface level (both physical devices and software apps in AllJoyn can expose interfaces that have members). Analysis of the SmartThings framework shows that apps are given access at the level of the device (not very fine grained) as a result of a flaw in the design. This leaves apps over-privileged and increases the risk of device exploitation. Over-privilege appears to be an issue whether it was by design or as a result of a flaw in the design. Good security strategies to follow should include:
\begin{enumerate}
    \item \textbf{The less unnecessary access the better.}
    \item \textbf{Informing the user in case of an exploitation can help mitigate the consequences.}
\end{enumerate}
It is easier for an entity to misbehave if it has more access than expected. In case this happens, it is necessary that the user is informed to mitigate the risks. 
It is worth to mention that mitigation actions may be visible or invisible to users.
For example, a visible action could be the unlocking of a door, while an invisible one could be transfer of the information electronically on the network. Zhang \emph{et al}. \cite{zhang2018homonit} proposed an intrusion detection system (IDS) to solve this problem in the SmartThings framework.  It sniffs the wireless communications in the smart home and raises an alarm when an anomaly is detected during runtime. This system relies on natural language processing (NLP) techniques of an application’s description. They achieved a 0.98 true positive rate. 
Another work by Tian \emph{et al.} attempts to include the user in what actually happens. 
They analyzed an app’s description, code and annotations to provide the user with better permission requests that are more inclusive of what the app actually uses. While this is useful to avoid over-privileged apps, the user could still benefit from tracking the app performance at runtime.
In this paper, we consider access control models that allow access at the level of the device, such as in SmartThings Smart-home platform.

Using the code fragment  Listing \ref{lst1} as an example, we illustrate over-privilege in SmartApps in further details. When the user permits a SmartApp to control a capability of a device, this device can be accessed through its declared id in the input statement. For example, if we look at the preferences section in Listing \ref{lst1}, when the user selects a device that supports the Motion Sensor capability, this device will be bound to the id \emph{themotion}. Using this id, commands and attributes of the capability can be accessed. However, some devices might support more than one capability. For example, the Phillips Hue Bulb supports both the Switch capability and the ColorControl capability \cite{writing_your_first_SmartApp}. Due to the flaw in the design, the id is actually a reference to the device as a whole. As a result, a SmartApp that is privileged to access a device for a specific capability, can access commands and attributes of all the capabilities in the bound Device Handler. This over-privilege in SmartApps is caused by coarse SmartApp-SmartDevice binding \cite{FernandesJP16}.

Another over-privilege case in the SmartThings framework is caused by coarse-grained capabilities \cite{FernandesJP16}. A coarse-grained capability provides the SmartApp with all the commands and attributes of the permitted capability, even if it only needs a subset of them. It is considered an over-privilege when the user permits the app to access a capability for the specified functionality, without knowing an app can control and access the entire capability.

Over-privilege in SmartApps occurs in two main scenarios: Acquiring more capabilities than permitted, and acquiring more commands and attributes than needed. For each scenario, we will provide a basic example of how over-privilege might occur in SmartApps. For fast proof-of-concept attacks, we used the web-based SmartThings IDE and simulated devices instead of physical ones.
\subsubsection{Over-Privilege Caused by Coarse SmartApp-SmartDevice Binding}
Once the SmartApp is authorized to access a specific capability of a device, the SmartApp can access all the capabilities implemented by the device handler (SmartDevice). To simulate this scenario, we need a device handler that supports at least two capabilities. For this example we used three capabilities: the Presence Sensor, the Lock and the Battery capabilities. We will write a SmartApp that unlocks the lock on arrival:
\vspace{3cm}
\begin{enumerate}
\item In the preferences method, request the Presence Sensor and the Battery capabilities: 
%\begin{minipage}{\linewidth}
\begin{lstlisting}[style=Groovy, caption={Over-privileged SmartApp, Case 1-step1}, label=list:CaseA1]
preferences 
{ section("Select devices")
  { input "thepresence", "capability.presenceSensor", title: "Select a presence sensor"
    input "thebattery", "capability.battery", title: "Select a battery" }}
 \end{lstlisting}
%\end{minipage}
\item Subscribe to the presence event handler in both installed() and updated() methods.
\item In the presence event handler, assuming \emph{thebattery} device supports the Lock capability as well, trigger the \emph{unlock} command and check the status of the lock through the \emph{currentLock} attribute:
%\begin{minipage}{\linewidth}
\begin{lstlisting}[style=Groovy, caption={Over-privileged SmartApp, Case 1-step2}, label=list:CaseA2]
def presenceHandler(evt)
{   if (thebattery.currentLock == "locked")
    {   thebattery.unlock()
        log.debug "Lock status: $thebattery.currentLock" }}
\end{lstlisting}
%\end{minipage}
\end{enumerate}
We run this SmartApp in the simulator with a device handler that implements the three capabilities, which simulates a device with those capabilities. We observed that \emph{thebattery} device successfully accessed the Lock capability of the device. This example proves that the \emph{thebattery} id is a reference to the whole device that is bound to the Battery capability, rather than to the Battery capability only. Thus, the SmartApp is over-privileged and has access to all the capabilities that this device supports.
\subsubsection{Over-Privilege Caused by Coarse-Grained Capabilities}%\label{Case3Intro}
A SmartApp always acquires all commands and attributes implemented by a permitted capability, even if it only needs a subset of them. This could be dangerous because different commands often have different levels of risk if the SmartApp is exploited \cite{FernandesJP16}. Typically, an app specifies its functionality in the description section, which often implies how it will use the capabilities. 

In Listing \ref{descr1}, the description of the application is: "Unlock the front door on arrival". As users, we understand that the action to be taken is the unlocking of the door, and it needs to happen after the person arrives to the place. When the person arrives, this indicates a change in the state of the presence.
%\begin{minipage}{\linewidth}
\begin{lstlisting}[style=Groovy, caption={SmartApp's Description in Free-Form Text}, label= descr1]
definition(
	name: "UnlockDoorApp",
	namespace: "",
	author: "",
	description: "Unlock the front door on my arrival",
	category: "",
	iconUrl: "",
	iconX2Url: ""
)
\end{lstlisting}
%\end{minipage}
Listing \ref{cap1} shows that the app requested two capabilities: the lock capability and the presenceSensor capability and their titles are shown to the user. The user can then deduce which commands and attributes are needed for this application to perform its functions. Specifically, the \emph{presence} attribute and the \emph{unlock} command only.
%\begin{minipage}{\linewidth}
%\vspace{5cm}
\begin{lstlisting}[style=Groovy, caption={SmartApp's Requested Capabilities}, label=cap1]
preferences
{   section("Select devices")
    { input "thepresence", "capability.presenceSensor", title: "Select a presence sensor"
      input "thelock", "capability.lock", title: "Select a lock"
    }
}
\end{lstlisting}
%\end{minipage}{}

If we look into the permission standards table in Listing \ref{cap2}, we notice that the lock capability has more commands and attributes than just the unlock command.

%\begin{minipage}{\linewidth}
\begin{lstlisting}[style=Groovy, caption={Segment from the Capabilities Table}, label=cap2]
light -> off
light -> on
lockOnly -> lock
lock -> lock
lock -> unlock
...
powerSource -> powerSource
presenceSensor -> presence
\end{lstlisting}
%\end{minipage}
We want to make sure that the application is not taking advantage and using non implied resources. To do this we search for the accessed commands and attributes in the application. In Listing \ref{list:CaseB2}, we have an example of the device \emph{thelock} actuating the \emph{lock} command to lock the door.
 
After running this SmartApp in the simulator, we find that the app did gain more commands than needed and could access them successfully.
This proves that the app is over-privileged and this is exactly what we want to detect when analyzing the applications.
\begin{table*}[t!]%*
\begin{center}
  \caption{Causes of Over-Privilege in SmartThings}
  \label{table:causes}
  \begin{adjustbox}{width=0.9\textwidth,totalheight={1\textheight},keepaspectratio}%
    \begin{tabular}{ccc}
  %  \toprule
    \multicolumn{1}{c}{\begin{tabular}[c]{@{}c@{}}Case1: Coarse SmartApp-\\ SmartDevice Binding\end{tabular}}     & Case2: Semantically Over-Privileged             & Case3: Coarse-Grained Capabilities        \\ 
  %  \midrule
    - Request set of capabilities A         & \begin{tabular}[c]{@{}l@{}}- Infer from free-form text the need \\of set of capabilities A\end{tabular}             & \begin{tabular}[c]{@{}l@{}}- Infer the use of a set of commands \&\\ attributes A, whose owner capability is\\ properly requested\end{tabular} \\
    \begin{tabular}[c]{@{}l@{}}- Access set of capabilities B,\\ such that A $\nsupseteq$ B\end{tabular} & \begin{tabular}[c]{@{}l@{}}- Request set of capabilities B,\\ such that A $\nsupseteq$ B\end{tabular} & \begin{tabular}[c]{@{}l@{}}- Access set of commands \& attributes B,\\ such that A $\nsupseteq$ B\end{tabular} \\
   % \bottomrule
  \end{tabular}
  \end{adjustbox}
\end{center}
\vspace{-0.5cm}
\end{table*}%*\subsection{Over-privilege in SmartApps}

\begin{figure}[b!]
\centerline{\includegraphics[width=.4\textwidth]{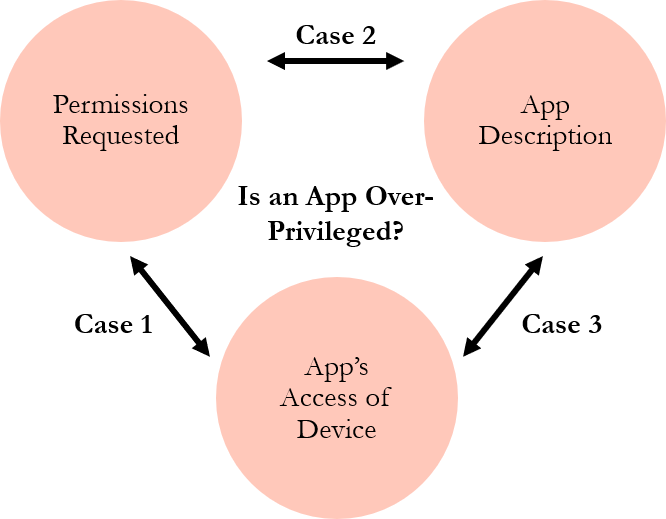}}
\caption{App Components Used in Analysis}
\label{fig:over_privilege_relationship}
\end{figure}
\begin{figure*}[t!]
\centering{
\includegraphics[width=0.65\textwidth]{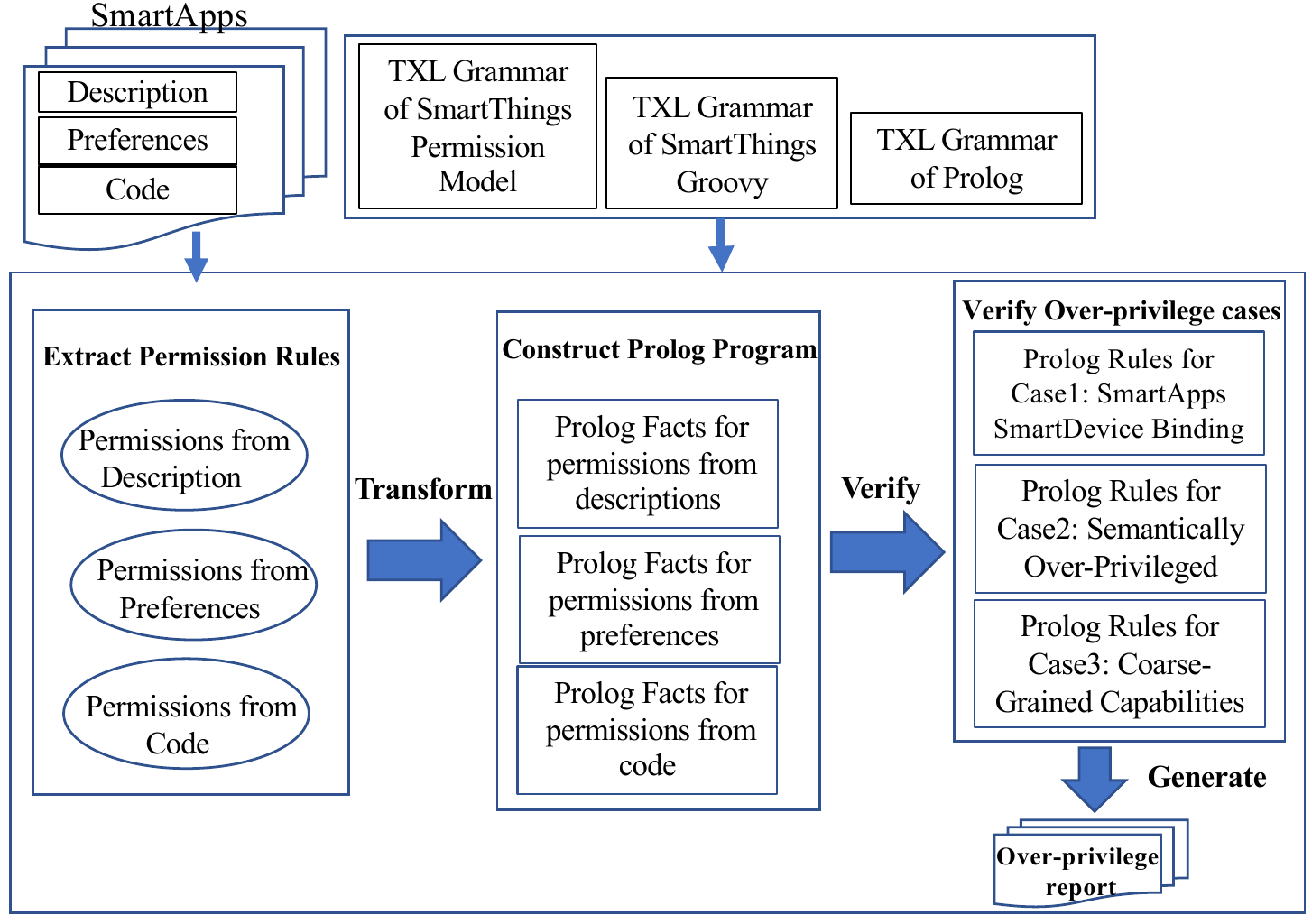}
\vspace{-0.1cm}
\caption{Approach Architecture}}
\vspace{-0.3cm}
\label{fig:approach_diagram_E}
\end{figure*}
%minipage is necessary to not confuse the reader
\begin{minipage}{\linewidth}
\begin{lstlisting}[style=Groovy, caption={In the presence event handler, after unlocking the lock, try to lock the device again}, label=list:CaseB2]
def presenceHandler(evt)
{   if (thelock.currentLock == "locked") 
    {   thelock.unlock()
        thelock.lock() }}
\end{lstlisting}
\end{minipage}

\begin{comment}
\begin{figure}[t!]
\centerline{\includegraphics[width=.2\textwidth]{Figures/inclusivity.png}}
\caption{Set of Capabilities A is a superset of B}
\label{fig:inclusivity}
\end{figure}
\end{comment}

To analyze an app for this case of over-privilege, we examine the inconsistencies between two sections of the app: the app's description and the app's code that accesses the device. Figure \ref{fig:over_privilege_relationship} illustrates the three types of over-privilege in the SmartThings framework based on inconsistencies between an app's components. Case 1 arises upon the usage of resources that are not requested by the app. While case 2 is the result of not mentioning the need for capabilities that are requested by the app. Case 3 results when the app describes the use of some of the commands and attributes of a requested capability but actually accesses the commands and attributes that are not needed according to the description.

Causes of over-privilege are summarized in Table \ref{table:causes}. In all three cases, if set A is not a super-set of set B (it is not inclusive of it), then it is considered to be an over-privileged app. In other words, set A has to include everything that is in set B to be considered a benign app. For example, in case 1, the app requests set of capabilities A, but some (or all) of the capabilities it accesses is from outside set A.
%See Figure \ref{fig:inclusivity}.
\section{Approach Overview}
Figure 2 %\ref{fig:approach_diagram_E} 
lays the three main components of our MDE approach used to verify over-privilege access in SmartThings Apps. 
This approach takes as input the SmartApp, TXL grammars of SmartThings permission model and SmartThings Groovy and a Prolog program containing the rules of over-privilege in SmartThings SmartApps. 
The first stage recovers the permission rules from the SmartThings app from the description, the preferences and the code. 
The recovered permission model conforms to a meta-model we constructed for SmartThings, and presented in Figure \ref{fig:metamodel}. 

Stage two generates the Prolog program and creates the executable from the Prolog main program which includes: the Prolog rules for each over-privilege case and the Prolog facts extracted from the SmartApp. 
The last step is to run the Prolog program to produce the final report with the over-privilege results. Each stage will be further discussed in the following sections.

\begin{comment}
\begin{figure}[p]
\centering
\includegraphics[width=1\textwidth]{Figures/extended_approach_diagram_2.png}
\caption{ChYP Extended Approach Diagram - Part 2}
\label{fig:extended_approach_diagram2}
\end{figure}
\end{comment}
\subsection{Extracting Permission Rules}
The process of extracting the permission model works by taking the smartapp as an input and extracting all the permission rules (if there is more than one). For example, in an IFTTT app there is one permission rule per app, but in a SmartThings app, it can be more complicated and include more than one \cite{alhanahnah2019advanced}. \hlc[highlight]{In order to automate the process we first need to define a meta-model that describes the permission model for the SmartThing platform.}
\subsubsection{Smart-Home Permission Meta-Model Definition}\label{sec:metamodel}
\begin{figure*}[t!]
\centering
\includegraphics[width=.65\textwidth]{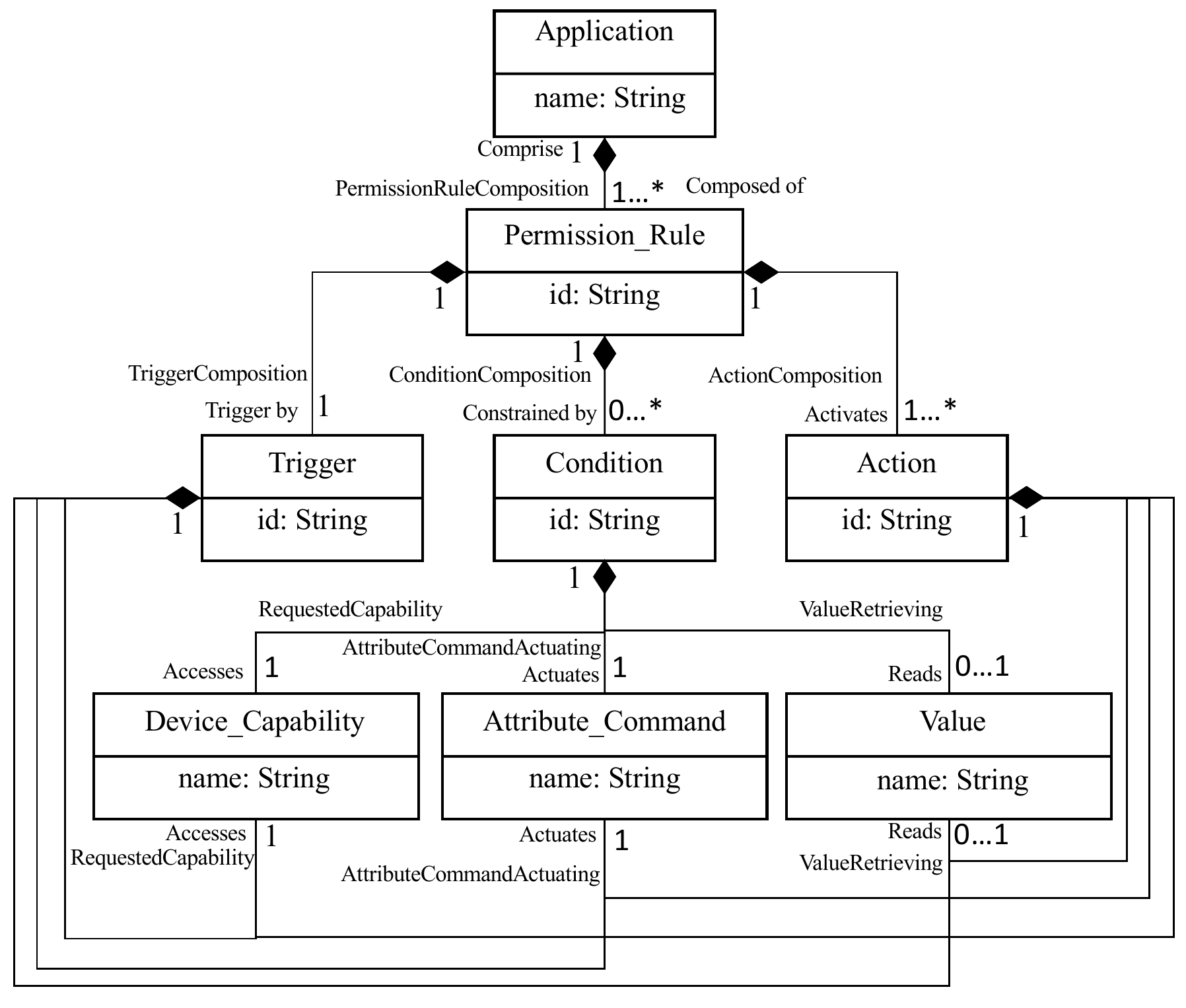}
\vspace{-.2cm}
\caption{Smart-Home Permission Meta-Model}
\vspace{-.5cm}
\label{fig:metamodel}
\end{figure*}

\hlc[highlight]{In this paper, we have developed a simplified meta-model that abstracts the permission model used in SmartThigs platform.} Figure~\ref{fig:metamodel} demonstrates the smart-home permission meta-model, which we adopted and revised from IoTCOM and SmartHomeML \cite{alhanahnah2019advanced,einarsson2017smarthomeml}. The Permission Rule in the meta-model (Figure \ref{fig:metamodel}) maps to the Behavioural Rule in IoTCOM. We adjusted the triggers in their model to be 1 trigger in ours, as each trigger leads to a specific set of conditions and actions. 
%This adjustment does not affect the overall extracted behaviour of the app, but it produces two permission rules instead of one in case they share the same conditions and actions. 
As for the conditions, we adjusted it to be zero or more, as not all rules have conditions. At last, if there exists no actions, then the permission rule is not taken into consideration.

We adopted from SmartHomeML the following:
\begin{itemize}
\item The “Query Action” translates to using the "attribute/command" and "value" with one of the rule components. In SmartThings it would typically either be a trigger or a condition. This is because one cannot change the value directly, but rather by actuating the command. 

\item The “Control Action” translates to using the "attribute/command", which in this case means command. In SmartThings it would typically be used in an action, not a trigger or a condition.

\item The "Skill (capability)" is the "device (capability)" in our model.
\end{itemize}

This concludes that our adjusted Smart-Home meta-model has the following structure: A permission rule consists of a trigger, a group of conditions (if any), and at least one action. Then, each trigger, condition and action has a device/capability, a command or an attribute and a value.
In SmartThings apps, a trigger is the entry point to the application. It triggers the associated event handler. 
This pair of trigger and event handler is declared in a subscribe statement, the example in Listing \ref{subscribe} shows an app subscribed to the \emph{`present'} event that will run the \emph{`presence'} event handler. 
\begin{lstlisting}[style=Groovy, caption={Trigger and Event Handler Subscription}, label=subscribe]
def initialize() {
	subscribe(driver, "presence.present", presence)}
\end{lstlisting}{}
The conditions are translated from the If statements in the event handlers, see example in Listing \ref{condition} where the condition \emph{If the event value is `present'} that will be translated to a condition in the permission rule. 

\begin{lstlisting}[style=Groovy, caption={If Statement to be Translated into a Condition}, label=condition]
def presence(evt) {
	if (evt.value == "present"){
    	... }}
\end{lstlisting}
The actions are anything else that is a direct access of the devices and resources. For example, in Listing \ref{action} the app actuates the \emph{`on'} command in the switch device.
\begin{lstlisting}[style=Groovy, caption={Non If Statement in Event Handler to be Translated into an Action}, label=action]
def turnLightsOn() {
    switches?.on()}
\end{lstlisting}
\subsubsection{Semantic Annotation and Preparation for Analysis}
\hlc[highlight]{The first challenge we encountered is how the approach will parse the natural language description presents in the app under analysis into the permission rule meta-model described in Figure }\ref{fig:metamodel}. For that, we used semantic annotation to properly annotate free-form text according to the permission meta-model we defined in the previous section.  The annotation relies on the extensive coverage of the grammar of words and phrases that map to objects of the meta-model \cite{annotating}. 

To automate the process of semantic annotation, we use TXL \cite{txl_doc1}, a hybrid functional/rule-based language with deep pattern matching that serves well for software structural analysis and source transformation.
To program in TXL, a context-free grammar of the input is defined (whether it's a programming language or something else). 
After that, transformation rules are provided that describe how the input will be transformed. 
The rules must follow the grammar to allow for a correct transformation.
\subsection*{\textbf{Generating TXL Grammar of SmartThings Permission Model}}
\hlc[highlight]{An essential part of the grammar needed for the analysis is the permission model of SmartThings. 
The permission model specifies the supported capabilities in SmartThings, in addition to the commands and attributes of each capability. 
A list of SmartThings capabilities can be found online.} \cite{capabilities}
\hlc[highlight]{Listing} \ref{capabilities_grammar}
\hlc[highlight]{Provides a snippet of the grammar built from the permission model taken from SmartThings documentation. 
We define the devices (capabilities) as in the Listing, as well as the commands and attributes and their values. 
Then, for each entity listed in the definitions, such as \emph{`accelerationSensor'} in the Listing, we define it according to the grammar inference process. (Section} \ref{grammar_inference})

\begin{lstlisting}[style=Groovy, caption={Grammar Snippet of SmartThings Permission Model}, label=capabilities_grammar]
% Definition of all capabilities
define device
    [accelerationSensor] | [alarm] | [audioNotification] | [battery] | [beacon] |
    [bulb] | [button] | [carbonDioxideMeasurement] | [colorControl] | [colorTemperature] |
    [configuration] | [consumable] | [contactSensor] | [doorControl] | [energyMeter] |
    [estimatedTimeOfArrival] | [garageDoorControl] | [holdableButton] | [illuminanceMeasurement] | 
    [imageCapture] | [indicator] | [infraredLevel] | [light] | [lockOnly] | [lock] | 
    [mediaController] | [momentary] | [motionSensor] | [musicPlayer] | [notification] | [outlet] | 
    [pHMeasurement] | [polling] | [powerMeter] | [powerSource] | [presenceSensor] | [refresh] | ...
end define

% Definition of the 'accelerationSensor' capability
define accelerationSensor
...
end define

% Definition of all commands
define command
...
end define

% Definition of all attributes
define attribute
...
end define

% Definition of all values
define value
...
end define
\end{lstlisting}
 \subsection*{\textbf{Grammar Inference}}\label{grammar_inference}
To build the needed grammar that enriches the annotations process, We opted for an approach inspired by grammar inference \cite{parekh2000grammar}. 
\hlc[highlight]{It is an important stage in building the grammar for this approach as it helps in capturing the semantics of the natural language statements in the apps.
To perform grammar inference, we used a public dataset.}
\footnote{\url{https://github.com/SmartThingsOverprivilege/smartthings_overprivilege_dataset}}
\hlc[highlight]{
We started with a subset of the apps, focusing on the apps natural language statements and manually tagging each word with a suitable entity mapped to the SmartThings permission model (if possible). Based on the tagging, we inferred the grammar terminal and non-terminal constructs and their relationships. We applied the first version of the grammar to the natural language statements we manually tagged and that to ensure the grammar is capable of producing the accurate annotation to the statements once they are parsed in the form of XML representation. If the annotation was successful, we repeat the same process on a new set of natural language statements from apps under analysis. The process of grammar inference is manual and tedious process as it requires expert's knowledge with mapping NL statements words to the SmartApps permission-system meta-model. The process needed multiple iterations until we were confident that the grammar produced is capable of automatically annotating all natural language statements used in the dataset under analysis. 
 Listing} \ref{grammar_inference_ex}
\hlc[highlight]{provides an example of the grammar inference process and the resulting grammar.
Line 1 presents the app's description in natural language: "Turn your lights on when motion is detected."
We tag each word in the description, if possible. Lines 5-10 provide each word with the entity we tagged it with, some of them have multiple tagging. For example, line 16 defines the word "when" as an indicator for a trigger.
Lines 14-38 explain how the tagging fits into the grammar we built.}

% for example, ...description > tagging

% Listing () shows the attribute \emph{`acceleration'} to be tagged with the words 'acceleration', 'sensed', and 'move' whenever encountered in natural language statements. 

%We build the grammar by learning the language and its syntax from real examples. 
%So, we manually built the grammar from natural language statements found in SmartApps.
%For example, Listing \ref{indicators} shows the inferred indicators for the permission rules, triggers and conditions. The word \emph{`when'} indicates that there is a trigger in this app. 
%All indicators have been inferred from the \hlc[highlight]{public} dataset. \footnote{\url{https://github.com/SmartThingsOverprivilege/smartthings_overprivilege_dataset}}

\begin{lstlisting}[style=Groovy, caption={Grammar Inference Example}, label=grammar_inference_ex]
App description: "Turn your lights on when motion is detected."

The manual tagging (the grammar inference process):

Turn > switch capability
lights > light capability
on > on attribute, command or value
when > indicator for trigger
motion > motion attribute or motionSensor capability
detected > active value

Snippet of the resulted grammar:

keys
    % indicators for trigger
    when whenever case
    ...
end keys

define switch
	'switch | 'turn
end define

define light
	'light | 'lights
end define

define motionSensor
	'motionSensor | 'motion
end define

define motionAttribute
	'motion
end define

define activeValue
	'active | 'detected
end define
\end{lstlisting}

\vspace{-0.3cm}
\subsection{Constructing Prolog Facts for SmartThings Permission-rules}
%\subsubsection{Mapping the Permission Meta-Model to Prolog Facts}
%The first problem we encountered is how the tool will parse the description into the permission rule meta-model in Figure \ref{fig:metamodel}. 
\hlc[highlight]{In order to successfully verify that the SmartApps do not have any type of over-privilege scenarios, the extracted permission artifacts needs to be transformed into a formal verification engine.  Using Prolog provides a flexible, scalable  and concise definition of verification goals. It enables the definition of domain-specific rules for validation. Prolog has a backtracking mechanism to test several variants of program flows for the verification goals. For our implementation, we used the established Prolog system SWI-Prolog . We have used Prolog in our previous work for access control verification of web applications} \cite{alalfi} \hlc[highlight]{and it proved to be efficient and scalable. IT was also recently used for embedded systems verification} \cite{prolog}.
\hlc[highlight]{In this meta-model, a SmartApp is composed of one or more permission rules. Each permission rule has the following components: a trigger, zero or more conditions and at least one action. 

For our analysis, we first need to define the mappings between the meta-model we defined in previous section and the corresponding Prolog facts. Each permission rule component accesses a capability and either a command or attribute and its value. }

%To parse the permission rules from the description, our approach is designed to perform semantic annotation of the description with attributes that map to %the meta-model in the form of Prolog facts.
Table \ref{table:metamodel_to_prolog_facts} presents the mapping between the meta-model described in Figure \ref{fig:metamodel} and their translation to Prolog facts.
For example, the \emph{`Permission\_Rule()'} Prolog fact consists of the \emph{`source'} (which is either from the code or description), the app's name and the id of the permission rule. Multiple permission rules can be extracted from the same app. Then each rule can have a trigger, conditions and actions, and each is linked through the RuleId. Each entity in the meta-model has a Prolog fact that defines its attributes. Operation \emph{`RequestedCapability'} also has a Prolog fact to record the list of capabilities the app requested.
%\begin{center}
\begin{table*}%{ p{.6\textwidth} | p{.4\textwidth} }
\caption{Permission Meta-Model Mapping to Prolog Facts}
\label{table:metamodel_to_prolog_facts}
\centering
\resizebox{0.8\textwidth}{!}{%
\begin{tabular}{|p{7cm}|p{8cm}|}
\hline
\textbf{Prolog Fact} & \textbf{Description} \\ \hline
Application(Source, AppName):                                                                                                                               &
Represents the name of the app being analyzed. Source of the fact is either 'description' or 'code'.       \\ \hline
Permission\_Rule(Source, AppName, RuleId):                                                                                                                  &
Represents a permission rule in an app.                                                                                                             \\ \hline
Trigger(Source, RuleId, TriggerId):                                                                                                                         &
Represents a trigger in a permission rule.                                                                                                                  \\ \hline
Action(Source, RuleId, ActionId):                                                                                                                           &
Represents an action in a permission rule.                                                                                                                  \\ \hline
Condition(Source, RuleId, ConditionId):                                                                                                                     &
Represents a condition in a permission rule.                                                                                                                \\ \hline
Attribute\_Command(Source, “Trigger/Condition/Action”+Id, AttributeCommandName):                                 &
Represents an attribute/command that belongs to a trigger, condition or an action in a permission rule.          \\ \hline
Device\_Capability(Source, “Trigger/Condition/Action”+Id, DeviceCapabilityName):                                 &
Represents a device/capability that belongs to a trigger, condition or an action in a permission rule.           \\ \hline
Value(Source, “Trigger/Condition/Action”+Id, ValueName):                                                                                                    &
Represents a value that belongs to a trigger, condition or an action in a permission rule.                       \\ \hline
TriggerComposition(Source, AppName, RuleId, TriggerId, DeviceCapabilityName, AttributeCommandName, ValueName):   &
Represents the details of a trigger of a permission rule.                                                                                                   \\ \hline
ConditionComposition(Source, AppName, RuleId, ConditionId, DeviceCapabilityName, AttributeCommandName, ValueName):   &
Represents the details of a condition of a permission rule.                                                                                                     \\ \hline
ActionComposition(Source, AppName, RuleId, ActionId, DeviceCapabilityName, AttributeCommandName, ValueName):         &
Represents the details of an action of a permission rule.                                                                                                       \\ \hline
Capability(DeviceCapabilityName):                                                                                                                               &
Represents the name of a device/capability.                                                                                                                     \\ \hline
AttributeCommandOf(DeviceCapabilityName, AttributeCommandName):                                                                                                 &
Represents the name of an attribute/command that belongs to a device/capability.                                     \\ \hline
ValueOf(AttributeCommandName, ValueName):                                                                                                                      &
Represents the name of a value that belongs to an attribute/command.                                                                                            \\ \hline
RequestedCapability(AppName, Capability):                                                                                                                       &
Represents a capability requested by an app.                                                                                                            \\ \hline
\end{tabular}%
}
\end{table*}

%To start the transformation, we perform semantic annotation of the description with attributes that map to the meta-model. 
%\section {Extract Permission-Rules}
\subsubsection{Extracting Permission-Rule Facts from Description}
This is the first main stage in our approach. The inputs to this stage are:
\begin{itemize}
    \item SmartApp to be analyzed
    \item TXL grammar of SmartThings permission rule
    \item TXL grammar of SmartThings groovy
    \item Extracted data-types of the devices in the SmartApp
\end{itemize}
TXL parses and annotates the descriptions in the SmartApp into the permission rule meta-model. 
Upon annotation, each permission rule extracted is translated into the corresponding Prolog facts from table \ref{table:metamodel_to_prolog_facts}. 
Facts are then saved in a file for later use. 
Listing \ref{desc} is an example of a description of a SmartApp.
%zigbeeLightFollowsMe
\begin{lstlisting}[style=groovy, caption={Description of a SmartApp}, label=desc]
description:"Turn your lights on when motion is detected.",
\end{lstlisting}{}

\begin{lstlisting}[style=Prolog-pygsty, caption={Intermediate Representation of Extracting Permission Rules From Description Using XML}, label=intermediate]
<program>
 <repeat_description>
  <description><repeat_permission_rule>
    <permission_rule>
     <repeat_rule_component>
      <rule_component><action><repeat_component_element>
         <component_element><device><*switch*>~Turn~</ *switch*></device></component_element>
         <component_element><word><id>your</id></word></component_element>
         <component_element><device><*colorControl*>~lights~</ *colorControl*></device></component_element>
         <component_element><attribute><*onAttribute*>~on~</ *onAttribute*></attribute></component_element>
         <empty/>
        </repeat_component_element>
       </action>
      </rule_component>
      <rule_component><trigger>
        <trigger_indicator>when</trigger_indicator>
        <repeat_component_element>
         <component_element><device><*motionSensor*>~motion~</ *motionSensor*></device></component_element>
         <component_element><word><id>is</id></word></component_element>
         <component_element><value><*detectedValue*>~detected~</ *detectedValue*></value></component_element>
...
\end{lstlisting}
This description is parsed by TXL into the intermediate representation of the grammar \hlc[highlight]{using XML, as in Listing} \ref{intermediate}, then TXL will again translate it to Prolog facts (see listing \ref{desc_facts}). 
\hlc[highlight]{The intermediate representation in Listing}  \ref{intermediate}
\hlc[highlight]{shows the mapping of each word in the description to its match in the grammar of SmartThings meta-model. For example, the word \emph{`Turn'} is annotated as the device (capability) \emph{`switch'} and the word \emph{motion} is annotated as the device (capability) \emph{`motionSensor'}.}
We mainly care about \textit{triggerComposition}, \textit{conditionComposition}, \textit{actionComposition} and \textit{device\_capability} from these facts, but the rest could be useful in future work, in this paper we only mention the main facts that are actually used in the Prolog rules.

\begin{lstlisting}[style=Prolog-pygsty, caption={Prolog Facts Extracted from Description}, label=desc_facts]
application(desc, AppName).

permission_rule(desc, AppName, rule1).
% Action facts
action(desc, rule1, action1).
device_capability(desc, action1, *switch*).
attribute_command(desc, action1, *on*).
value(desc, action1, *on*).
actionComposition(desc, AppName, rule1, action1, *switch*, *on*, *on*).
% Trigger facts
trigger(desc, rule1, trigger1).
device_capability(desc, trigger1, ~motionSensor~).
attribute_command(desc, trigger1, ~motion~).
value(desc, trigger1, ~detected~).
triggerComposition(desc, AppName, rule1, trigger1, ~motionSensor~, ~motion~, ~detected~).
\end{lstlisting}
\subsubsection{Extracting Permission-Rule Facts from Preferences}
The input to this stage is the SmartApps and the grammars as in the first stage above. TXL analyzes the SmartApp for capabilities requested in the preferences section. Listing \ref{preferences} is an example of a SmartApp requesting the switch capability. This input statement is translated to a Prolog fact (as in listing \ref{requested}), and saved to the same facts file used previously. 
\begin{lstlisting}[style=Groovy, caption={Capabilities Requested by SmartApp}, label=preferences]
preferences {
	section("When I reach home, turn on the lights.") {
		input "switches", *"capability.switch"*, multiple: true }
}
\end{lstlisting}
\begin{lstlisting}[style=Prolog-pygsty, caption={Prolog Fact for Case 2}, label=requested]
requestedCapability(AppName, *switch*).
\end{lstlisting}{}
\subsubsection{Extracting Permission-Rule Facts from Code}
The third stage is to analyze the app code to extract facts for cases 1 and 3. 
The analysis extracts the permission rules from the code.
\hlc[highlight]{We start with the trigger, which is extracted from the subscribe methods in the app.}
Each subscribe method declares the event handler (or method) that will be executed upon the trigger. 
The event handler is analyzed to extract the conditions (any if statements), while all other statements are considered actions.

In listing \ref{trigger_event}, the trigger is the change to the motion sensor\hlc[highlight]{, specifically when "motion" attribute changes its value to "active".} This triggers the execution of the event handler \textit{motionActiveHandler}. The resulted trigger fact is shown in listing \ref{trigger_fact}. After that, the tool analyzes the code of the event handler for conditions and actions. The resulted two facts would typically be as in listing \ref{condition_action_facts}. 
In the case that we extracted either an attribute/command or a value, and this extracted word is considered both an attribute/command and a value in the TXL grammar, we fill it in both placed in the Prolog fact. 
This is because when we analyze the description of the app, it is hard to infer if this word should be filled in the attribute/command or the value in the Prolog fact.

\begin{lstlisting}[style=Groovy, caption={Trigger and Associated Event Handler}, label=trigger_event]
def installed()
{   subscribe(motion1, ~"motion.active"~, motionActiveHandler)}

def motionActiveHandler(evt) {
    if(switch1.currentValue(*"switch"*) == *"off"*){
        `switch1.on()`}}
\end{lstlisting}
\begin{lstlisting}[style=Prolog-pygsty, caption={Trigger Fact}, label=trigger_fact]
triggerComposition(code, AppName, rule1, trigger1, ~motionSensor~, ~motion~, ~active~).
\end{lstlisting}
\begin{lstlisting}[style=Prolog-pygsty, caption={Condition and Action Facts}, label=condition_action_facts]
conditionComposition(code, AppName, rule1, condition1, *switch*, *off*, *off*).
actionComposition(code, AppName, rule1, action1, `switch`, `on`, `on`).
\end{lstlisting}{}

%\subsection{Constructing and Running Prolog Model Checker}
\vspace{-0.3cm}
\subsection{Verifying Over-privilege cases}
\hlc[highlight]{We create the Prolog model checker from within the TXL main program of the tool. Line 1 in Listing} \ref{prolog_model_checker}
\hlc[highlight]{is the command used for the creation of the Prolog executable.
Our Prolog model checker takes three sources of information to carry out the analysis.
Table} \ref{table:var_def}
\hlc[highlight]{ defines the use and contents of each variable/file.
The model checker first takes the main program needed to specify which rules to run. It also needs the Prolog rules file containing the three Prolog rules for checking the three cases of over-privilege explained in the following sections. Lastly, we provide the Prolog facts file containing all the facts we extracted from the app's description, preferences and code.
After we have created the Prolog model checker, we pass the program's file name to the 'run' command to run the analysis and produce the analysis report.}

\begin{lstlisting}[style=Groovy, caption={Command to Create the Prolog Model Checker}, label=prolog_model_checker]
    construct createPrologExe [stringlit]
        _ [+ "swipl --goal=main --stand_alone=true -q -o "] [+ *prologExecutableFileName*] [+ " -c "] [+ *prologMainFile*] [+ *prologRulesFile*] [+ *prologFactsFile*]
    construct runPrologProgram [stringlit]
        _ [+ "run "] [+ *prologExecutableFileName*] [+ ".exe"]
    construct workHere [stringlit]
        _   [~system createPrologExe~]
            [~system runPrologExe~]
\end{lstlisting}

\begin{table*}[t!]
\centering
    \caption{\hlc[highlight]{Variables Definitions}}
    \label{table:var_def}
\resizebox{0.7\textwidth}{!}{%

\begin{tabular}{cc}
    \toprule
    Variable                 & Purpose                                                                                              \\
    \midrule
    prologExecutableFileName & \begin{tabular}[c]{@{}c@{}}The name of the output Prolog program that will be used to run \\ the model checker\end{tabular}                        \\
    prologMainFile           & The main Prolog file used to initiate the Prolog checker                                                                                        \\
    prologRulesFile          & \begin{tabular}[c]{@{}c@{}}The Prolog file containing the three main rules for detection of \\ over-privilege cases\end{tabular}                \\
    prologFactsFile          & \begin{tabular}[c]{@{}c@{}}The Prolog file containing the Prolog facts extracted from the \\ description, preferences and the code\end{tabular}    \\
    \bottomrule
\end{tabular}}
\end{table*}

\subsubsection{Detection of Case 1: Over-Privilege Caused by Coarse SmartApp-SmartDevice Binding}
\subsection*{\textbf{Prolog Rules for case 1:}} This case of over-privilege occurs when there is an inconsistency between the \textit{Permissions Requested} and \textit{App's Access of Device}. To check for occurrences of over-privilege case 1, we need to inspect if any of the devices approved for this app and has more than one capability, was misused to access a capability not intended for this app. A query to the Prolog rule needs to satisfy the following:
\begin{itemize}
    \item a resource (a command or a value of an attribute), has been used in a trigger, a condition or an action
    \item this resource is not a command of the capability in this trigger, condition or action
    \item this resource is not a value of an attribute of the capability in this trigger, condition or action
    \item this resource belongs to another capability in the SmartThings capability model
    \item the capability and the resource are both non empty in the facts
\end{itemize}

\hlc[highlight]{A complete Prolog rule for case 1 can be found in Appendix A.}

\subsection*{\textbf{Prolog Facts for case 1}}
Given the facts in the listing below, an over-privilege of case 1 must be reported. The fact on line 13 states that the device \textit{accelerationSensor} was used to actuate the \textit{on} command, which is a command of another capability and not of \textit{accelerationSensor}'s.
%\begin{minipage}{\linewidth}
\begin{lstlisting}[style=Prolog-pygsty, caption={Sample Facts Relevant to Over-Privilege Case 1}, label=prolog_facts_Case1]
/* SmartThings Capability model facts */
capability(~accelerationSensor~).
attributeCommandOf(~accelerationSensor~,~acceleration~).
valueOf(~acceleration~,*active*).
valueOf(~acceleration~,*inactive*).
capability(switch).
attributeCommandOf(switch,switch).
attributeCommandOf(switch,off).
attributeCommandOf(switch,on).
valueOf(switch,off).
valueOf(switch,on).
/* Fact used to check case 1 */
triggerComposition(code,AppName,ruleNumber,trigger1,~accelerationSensor~,`on`,na).
\end{lstlisting}
%\end{minipage}
\subsubsection{Detection of Case 2: Over-Privilege Caused Semantically}\label{Case2Intro}
\subsection*{\textbf{Prolog Rules for case 2}}
This case of over-privilege occurs when there is an inconsistency between the \textit{App Description} and \textit{Permissions Requested}. To detect over-privilege case 2, we need to inspect if the application declared or indicated in the description what capabilities it will be accessing. A query to the Prolog rule needs to satisfy the following:
\begin{itemize}
    \item the app have requested a certain capability (fact name: \textit{requestedCapability})
    \item there is no mention of this capability in the description facts (fact name: \textit{device\_capability})
    \item capability has a name and is not empty (na)
\end{itemize}
This listing below is the resulting Prolog rule to detect this case of over-privilege:
\begin{lstlisting}[style=Prolog-pygsty, caption={Prolog Rules to Detect Case 2 of Over-Privilege in SmartApps}, label=prolog_rules_Case2]
overprivilegedCase2(case2,AppName,Capability):-
    requestedCapability(AppName,Capability),
    not(device_capability(desc,_,Capability)),
    not(Capability=na).
\end{lstlisting}
\vspace{-0.5cm}
\subsection*{\textbf{Prolog Facts for case 2}}
Given the facts in the listing below, an over-privilege of case 2 must be reported. As non of the capabilities declared in the description facts is the \textit{switch} capability.
\begin{lstlisting}[style=Prolog-pygsty, caption={Sample Facts Relevant to Over-Privilege Case 2}, label=prolog_facts_Case2]
device_capability(desc,_,~presenceSensor~).
device_capability(desc,_,*door*).
requestedCapability(AppName, ~presenceSensor~).
requestedCapability(AppName, *door*).
requestedCapability(AppName, `switch`).
\end{lstlisting}
\vspace{-0.3cm}
\subsubsection{Detection of Case 3: Over-Privilege Caused by Coarse-Grained Capabilities}%\label{Case3Intro}
\subsection*{\textbf{Prolog Rules for case 3}}
This case of over-privilege occurs when there is an inconsistency between the \textit{App Description} and \textit{App's Access of Device}. It can happen as a result of coarse-grained capabilities. It means that a permitted capability is fully available to the app to access, with all its commands and attributes. This is considered an over-privilege because risk associated with giving access to different commands and attributes is not on the same level \cite{FernandesJP16}. For example, giving access to turn off the oven and have it be misused might be annoying, while misusing the turn on operation is very risky.

A query to the Prolog rule needs to satisfy one of the following:
\begin{itemize}
    \item The app performs an action in the code that is not mentioned in the description.
    \item There exists a combination of a trigger and an action in code, that is not mentioned in the description with any combination of the capability, attribute/command and the value. For example, suppose that there exists in the code a trigger (if presence changes), turn on lights (action), but, the description only declared the lights to turn off if you leave the house. There would be inconsistency between the facts in the description and the code, leading to reporting an over-privilege of case 3 (sample facts for this example are shown in listing \ref{prolog_facts_Case3}).
    \item Same case as the one above, but with differences in the combination of a condition and an action, rather than a trigger and an action. An example for this would be if we found in the code a condition that if the presence changes, it would turn the lights on, while in the description, it was declared that the lights should be turned on only after 10 AM. 
\end{itemize}

This listing below shows part of the resulting Prolog rule to detect this case of over-privilege:
\begin{lstlisting}[style=Prolog-pygsty, caption={Sample Prolog Rules to Detect Case 3 of Over-Privilege in SmartApps}, label=prolog_rules_Case3]
overprivilegedCase3(case3,AppName,RuleIdCode,TriggerIdCode,ConditionIdCode,ActionIdCode,Capability,AttributeCommand,Value):-
    /*The whole action is missing in description*/
    *actionNotInDesc*(AppName,RuleIdCode,ActionIdCode,Capability,AttributeCommand,Value);
*actionNotInDesc*(AppName,RuleIdCode,ActionIdCode,Capability,AttributeCommand,Value):-
    actionComposition(~code~,AppName,RuleIdCode,ActionIdCode,Capability,AttributeCommand,Value),
    not(actionComposition(~desc~,AppName,_,_,Capability,AttributeCommand,Value)),
    not(*threeNAs*(Capability,AttributeCommand,Value)),
    not(*actionTwoNAs*(AppName,Capability,AttributeCommand,Value)).
%* Capability, 
*threeNAs*(Capability,AttributeCommand,Value):-
    *Capability=na*,
    *AttributeCommand=na*,
    *Value=na*.
/*  Use Case:
    Desc: "Lock the front door"
    Code: "Unlocks the door" */
*actionTwoNAs*(AppName,Capability,AttributeCommand,Value):-
    *AttributeCommand = na*,
    *Value = na*,
    actionComposition(desc,AppName,_,_,Capability,_,_).
\end{lstlisting}

\subsection*{\textbf{Prolog Facts for case 3}}
Given the facts in the listing below, an over-privilege of case 3 must be reported. As there is an inconsistency between the facts of the permission rules in the description with the ones in the code.
%\begin{minipage}{\linewidth}
\begin{lstlisting}[style=Prolog-pygsty, caption={Sample Facts Relevant to Over-Privilege Case 3}, label=prolog_facts_Case3]
/* Facts in description */
permission_rule(desc, AppName, ruleNumber).
triggerComposition(desc, AppName, ruleNumber, trigger1, ~presenceSensor~, ~presence~, ~notpresent~).
actionComposition(desc, AppName, ruleNumber, action1, *switch*, ~off~, ~off~).
/* Facts in code */
permission_rule(code, AppName, ruleNumber).
triggerComposition(code, AppName, ruleNumber, trigger1, ~presenceSensor~, ~presence~, na).
actionComposition(code, AppName, ruleNumber, action1, *switch*, `on`, `on`).
\end{lstlisting}
%\end{minipage}

\section{Evaluation}
In this section, we \hlc[highlight]{evaluate our MDE approach and tool designed to detect the three cases of over-privilege.} 
%The approach is implemented in our tool. %called ChYP (Check Your Privilege). 
The implementation makes use of the TXL and Prolog languages. \hlc[highlight]{TXL} grammars \hlc[highlight]{are needed for this tool implementation, including grammars of:} SmartThings groovy, SmartThings permission model, Prolog facts and an XML grammar from the TXL project \cite{TXL_World} as a helper grammar in the transformation.
%After that, we will define some evaluation measures and conduct manual validation of our results. 
\hlc[highlight]{To evaluate our approach, we obtained a dataset from publicly available sources, these include: SmartThings marketplace, SmartThings community and SmartThings forums. Those apps are developed by third-party developers and their source code is publicly available, this makes the dataset a suitable target for the evaluation. We also used the dataset provided by Zhang \emph{et al.}} \cite{zhang2018homonit} \hlc[highlight]{to evaluate the detection of case 3. We refer to this dataset as "HoMonit" in this evaluation.}
% (the acquiring of the dataset is discussed in section} \ref{sec:dataset}).
\hlc[highlight]{We evaluate the effectiveness of our approach by answering the following research questions:}
\begin{enumerate}
    \item \hlc[highlight]{RQ1: What is the performance of our approach? How scalable is our analysis when applied to a large dataset of SmartThings apps? Can our automated analysis report the 3 cases of over-privilege in SmartThings apps?} \hlc[highlight] {(Section} \ref{sec:rq1})
    \item \hlc[highlight]{RQ2: How does privilege escalation detection performed by ChYP} \cite{chyp} \hlc[highlight]{compare to the MDE approach presented in this paper?} \hlc[highlight]{(Section}\ref{sec:rq3})
    \item \hlc[highlight]{RQ3: How effective is our tool in detecting privilege escalation in SmartThings apps when evaluated for precision and recall?}%. What is the accuracy of our tool that employs MDE for the model extraction and Prolog for the model checking?} 
    \hlc[highlight]{ (Section} \ref{sec:rq2})
       % \item \hlc[highlight]{RQ4: What is the performance of our tool? How scalable is our tool in performing analysis on a large dataset of SmartThings apps?} \hlc[highlight]{(Section} \ref{sec:rq4})
\end{enumerate}
\begin{comment}
\subsection{Dataset}\label{sec:dataset}
To evaluate our tool and its effectiveness of detecting over-privilege in SmartThings apps, we used multiple datasets.
The source of the datasets are from the SmartThings marketplace, SmartThings community and SmartThings forums. Those apps are developed by third-party developers and their source code is publicly available, this makes the dataset a suitable target for the evaluation.
We also used the dataset provided by Zhang \emph{et al.} \cite{zhang2018homonit} to evaluate the detection of case 3. We refer to this dataset as "HoMonit" in this evaluation.

to compare results with HoMonit. 
This dataset consisted of 30 benign apps, that were modified to produce another 30 malicious apps. 
Each of the 30 apps were tested either on ZigBee or Z-Wave communication protocols. The dataset also has over-privilege and event-spoofing vulnerabilities. 
Since our approach does not depend on the communication protocol, or detects event-spoofing, we eliminated the apps that did not affect our experiment. 
After the elimination, we had 14 apps out of 30 for each of the benign and malicious datasets.
\end{comment}

\subsection{Automated Analysis Results}\label{sec:rq1}
%We evaluate our tool in detecting over-privilege in the dataset we created. 
%We evaluate our experimental results by manually checking the automated detection results, then calculating the precision and recall. 

\begin{figure*}[t!]
    \centering
    \caption{\hlc[highlight]{Distribution of Over-Privilege Cases Across the Dataset}}
    \label{fig:pie}
    \includegraphics[width=0.5\textwidth]{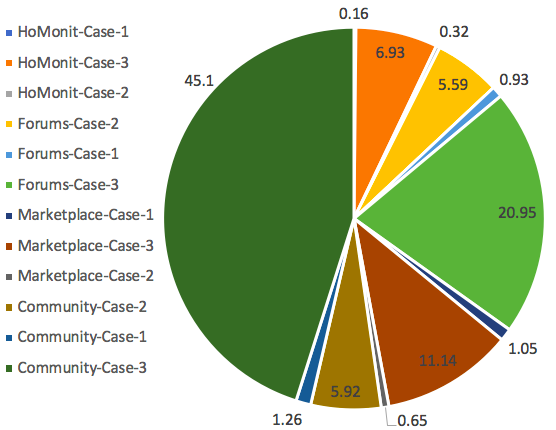}
\end{figure*}

\hlc[highlight]{In this section we address RQ1. 
First, we run the tool on the publicly available dataset.}
\hlc[highlight]{We then observe the execution of the tool on the dataset whether it completed successfully and examine its coverage in reporting the three over-privilege cases.} 
Table \ref{table:initial_running_resutls} details the statistical over-privilege results obtained from running our tool on the dataset.
\hlc[highlight]{The tool ran successfully and completed all stages of the analysis.
A total of 230 apps were analysed and} over 2000 occurrences of over-privilege were reported across the dataset from cases 1, 2 and 3. 
\hlc[highlight]{Figure} \ref{fig:pie}.
\hlc[highlight]{illustrates the distribution of over-privilege cases in each of the dataset classifications derived from Table} \ref{table:initial_running_resutls}.
\hlc[highlight]{In all four classifications, case 3 dominated the reported over-privilege occurrences. 
This is reasonable because it is the easiest case to fall into. In case 3, the capability is correctly requested and it is not used to access commands from another capability. 
What happens in case 3 is that the developer ambiguously declares how or when the commands and attributes of the capability will be accessed (whether accidentally or intentionally).
So, even if for example the developer declared that the \emph{`lock'} and \emph{`unlock'} commands of the lock capability will be actuated on arrival and take off but did not specify which command is paired with which event, this might be suspicious to the user and should be marked as over-privilege.}

%Results show that our tool reaches 78.12\% precision and 92.59 recall for detecting case 1,  88.24\% precision and 78.95\% recall for detecting case 2, and 70.37\% precision and 95\% recall for detecting case 3. 
%We also had benign dataset for case 2 and 3. Our tool achieved 100\% precision and 82.76\% recall for case 2 and 100\% precision and 39.22\% recall for case 3. 
%Causes of false positives are explained in section \ref{futurework}.

\begin{table*}[t!]
\centering
  \caption{MDE-ChYP results on publicly available datasets}
  \label{table:initial_running_resutls}
  \begin{tabular}{cccccc}
    \toprule
    Dataset     & Total Apps & \begin{tabular}[c]{@{}c@{}} Total Cases \\ Reported \end{tabular} & Case 1 & Case 2 & Case 3 \\ 
    \midrule
    Homonit     & 19         & \hlc[highlight]{183}         & 4     & \hlc[highlight]{8}       & \hlc[highlight]{171}  \\
    Forums      & 50         & \hlc[highlight]{678}         & 23    & \hlc[highlight]{138}     & \hlc[highlight]{517}  \\
    Marketplace & 19         & \hlc[highlight]{317}         & 26     & \hlc[highlight]{16}      & \hlc[highlight]{275}  \\
    Community   & 142        & \hlc[highlight]{1290}        & 31     & \hlc[highlight]{146}     & \hlc[highlight]{1113} \\
    Total       & 230        & \hlc[highlight]{2468}        & 84     & \hlc[highlight]{308}     & \hlc[highlight]{2076} \\ 
  \bottomrule
\end{tabular}
\end{table*}

% homonit case 1 0.16%
% homonit case 2 0.32%
% homonit case 3 6.93%
% forums case 1 0.93%
% forums case 2 5.59%
% forums case 3 20.95%
% marketplace case 1 1.05%
% marketplace case 2 0.65%
% marketplace case 3 11.14%
% community case 1 1.26%
% community case 2 5.92%
% community case 3 45.10%

\subsubsection{\hlc[highlight]{Performance of the Tool}}\label{sec:rq4}
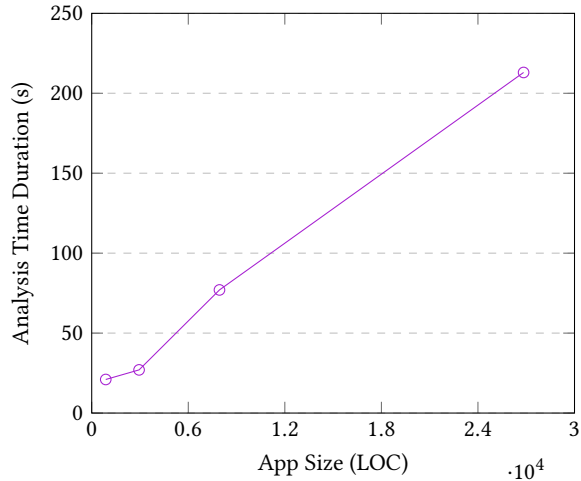
\begin{figure}
    \centering
    \caption{Analysis time of privilege escalation detection in SmartApps in relation to size of the dataset}
    %\label{fig:my_label}
    \label{fig:performance1}
\begin{tikzpicture}
\begin{axis}[
    xlabel={App Size (LOC)},
    ylabel={Analysis Time Duration (s)},
    xmin=0, xmax=30000,
    ymin=0, ymax=250,
    xtick={0,6000,12000,18000,24000,30000},
    ytick={0,50,100,150,200,250},
    %legend pos=north west,
    ymajorgrids=true,
    grid style=dashed,
]

\addplot[
    color=purple,
    mark=o,
    ]
    coordinates {
    (872,21)(2943,27)(7939,77)(26836,213)
    };
    \legend{}
    
\end{axis}
\end{tikzpicture}
\end{figure}

\begin{figure}
    \centering
    \caption{Analysis time of privilege escalation detection in SmartApps in relation to no. of Prolog facts}
    \label{fig:performance2}
\begin{tikzpicture}
\begin{axis}[
    xlabel={Total Prolog Facts},
    ylabel={Analysis Time Duration (s)},
    xmin=0, xmax=20000,
    ymin=0, ymax=250,
    xtick={0,4000,8000,12000,16000,20000},
    ytick={0,50,100,150,200,250},
    %legend pos=north west,
    ymajorgrids=true,
    grid style=dashed,
]

\addplot[
    color=purple,
    mark=o,
    ]
    coordinates {
    (1439,21)(2559,27)(7338,77)(19360,213)
    };
    \legend{}
    
\end{axis}
\end{tikzpicture}
\end{figure}
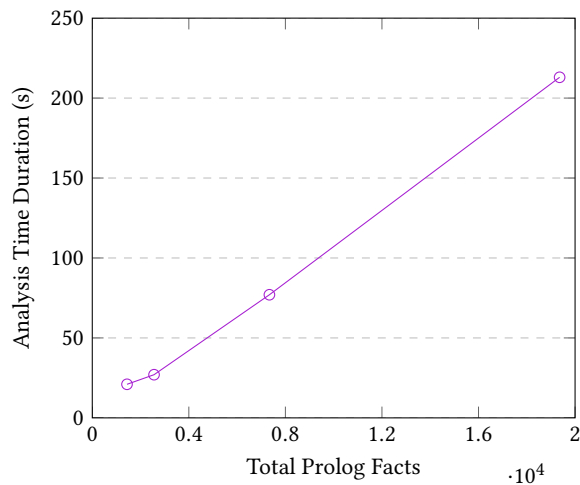

\hlc[highlight]{To answer RQ1 in regard to performance of the tool, we measure the analysis time for each dataset classification separately, as in Table} \ref{table:performance}. 
\hlc[highlight]{We ran the analysis on a PC with an Intel Core i7 2.2 GHz CPU processor and 20 GB RAM.
The data we gather to calculate the performance are the lines of code (LOC) for the SmartThings apps, the number of Prolog facts generated by each app and time spent in analysis.
Table } \ref{table:performance}
\hlc[highlight]{ details the performance data reported after running the evaluation experiment.
The largest bundle of apps, the Community dataset including 142 apps with more than 26,000 LOC, took a little over 3 minutes and a half to complete the analysis.
Figures} \ref{fig:performance1} 
\hlc[highlight]{and} \ref{fig:performance2} 
\hlc[highlight]{illustrate the time growth of the analysis in regards to the LOC and Prolog facts respectively.}
\hlc[highlight]{Both figures imply that the analysis time seems to be scaling linearly with the LOC of TXL and Prolog facts.}

\begin{table*}[t!]
\centering
\caption{\hlc[highlight]{Performance Statistics}}
\label{table:performance}
\begin{tabular}{ccccc}
    \toprule
    Dataset    & Total Apps & Total LOC & Total Prolog Facts & Analysis Time Duration \\
    \midrule
    HoMonit     & 19         & 872       & 1439               & 0m21s                \\
    Forums      & 50         & 7939      & 7338               & 1m17s             \\
    Marketplace & 19         & 2943      & 2559               & 0m27s                \\
    Commnunity  & 142        & 26,836    & 19,360             & 3m33s              \\
    \bottomrule
\end{tabular}
\end{table*}

\subsection{Comparison between the MDE-based Approach (MDE-ChYP) and the Static Analysis Approach (Chyp)} \label{sec:rq3}
\hlc[highlight]{To answer RQ2, we compare our MDE-based approach with our previous static analysis-based approach, ChYP} \cite{ours}, \hlc[highlight]{on two aspects: accuracy and coverage of privilege escalation detection.
The static analysis approach analyzes the syntax only and performs well in tracing complex functions and loops. 
The MDE approach provides a combination of static analysis, text processing, and logic-based reasoning.}
%The MDE approach analyzes the free-form text, and treats the code as free-form text as well.}

\begin{table*}[t!]
\centering
\caption{Detailed results of evaluation comparison between MDE-ChYP and the static analysis tool ChYP}
\label{table:comparison_detailed}
\begin{tabular}{ccccccccc}
\toprule
\multirow{2}{*}{Dataset} & \multirow{2}{*}{App}                  & \multirow{2}{*}{Case 1 Occurrences} & \multicolumn{3}{c}{ChYP} & \multicolumn{3}{c}{MDE-ChYP} \\
                         &                                       &                                    & TP      & FP      & FN     & TP      & FP      & FN     \\
\midrule
Homonit                  & zigbeeFloodAlert                      & 0                                  & 0       & 0       & 0      & 0       & 1       & 0      \\
Marketplace              & color\_coordinator                    & 6                                  & 6       & 0       & 0      & 6       & 0       & 0      \\
                         & enhanced\_auto\_lock\_door            & 0                                  & 0       & 0       & 0      & 0       & 1       & 0      \\
                         & keep\_me\_cozy                        & 1                                  & 1       & 0       & 0      & 1       & 0       & 0      \\
Forums                   & alarmThing\_AlertAll                  & 2                                  & 0       & 0       & 2      & 2       & 0       & 0      \\
                         & buffered\_event\_sender               & 1                                  & 0       & 0       & 1      & 1       & 0       & 0      \\
                         & fireCO2Alarm                          & 5                                  & 3       & 0       & 2      & 2       & 0       & 3      \\
                         & garage\_switch                        & 2                                  & 1       & 0       & 1      & 2       & 0       & 0      \\
                         & groveStreams                          & 1                                  & 0       & 0       & 1      & 1       & 0       & 0      \\
                         & initial\_state\_event\_sender         & 1                                  & 0       & 0       & 1      & 1       & 0       & 0      \\
                         & initialstate\_smart\_app\_v1\_2\_0    & 1                                  & 0       & 0       & 1      & 1       & 0       & 0      \\
                         & unbuffered\_event\_sender             & 1                                  & 0       & 0       & 1      & 1       & 0       & 0      \\
                         & zwave\_indicator\_manager             & 4                                  & 4       & 0       & 0      & 4       & 1       & 0      \\
Community                & flood\_alert                          & 0                                  & 0       & 0       & 0      & 0       & 1       & 0      \\
                         & garage\_door\_monitor                 & 1                                  & 0       & 0       & 1      & 1       & 0       & 0      \\
                         & initial\_state\_event\_streamer       & 1                                  & 0       & 0       & 1      & 1       & 0       & 0      \\
                         & keep\_me\_cozy\_ii                    & 1                                  & 1       & 0       & 0      & 1       & 0       & 0      \\
                         & lock\_it\_when\_i\_leave              & 0                                  & 0       & 0       & 0      & 0       & 1       & 0      \\
                         & medicine\_management\_contact\_sensor & 3                                  & 3       & 0       & 0      & 1       & 0       & 2      \\
                         & medicine\_management\_temp\_motion    & 3                                  & 3       & 0       & 0      & 1       & 0       & 2      \\
                         & ridiculously\_automated\_garage\_door & 0                                  & 0       & 0       & 0      & 0       & 2       & 0      \\
                         & smartblock\_linker                    & 2                                  & 0       & 0       & 2      & 2       & 0       & 0      \\
                         & step\_notifier                        & 3                                  & 3       & 0       & 0      & 2       & 0       & 1      \\
                         & the\_big\_switch                      & 1                                  & 0       & 0       & 1      & 1       & 0       & 0      \\
                         & Thermostats                           & 1                                  & 0       & 0       & 1      & 1       & 0       & 0      \\
Total                    & 25 Apps                               & 41                                 & 25      & 0       & 16     & 33      & 7       & 8    \\ \bottomrule
\end{tabular}
\end{table*}
\hlc[highlight]{For this comparison, we randomly picked 25 apps that reported occurrences of case 1 when we ran the MDE-ChYP on the public datasets in table} \ref{table:initial_running_resutls}.
\hlc[highlight]{The list of apps used in this experiments are displayed in table}
\ref{table:comparison_detailed}. 
\hlc[highlight]{We manually checked the apps for actual occurrences of case 1 and found 41 occurrences across the apps.
We ran both tools on this dataset of 25 apps, then we manually validated the results reported by both tools.
Through the validation of results from the static analysis based tool, we found 25 true positives and 16 false negatives.
While for MDE-ChYP tool, we found 33 true positives, 7 false positives and 8 false negatives. (see Table}
\ref{table:comparison_detailed}).

\hlc[highlight]{Table} \ref{table:comparison}
\hlc[highlight]{ displays the precision and recall of evaluating both tools on this selected dataset of 25 apps.}

Precision evaluates the tool in not returning non-over-privileged cases. 
While recall evaluates the tool in returning the over-privileged cases if present in an app \cite{annotating}. 
Below are the formulas used to calculate the precision and recall \cite{zhang2018homonit, annotating}. 
Precision is calculated by dividing the true positives (TP) by the true positives and false positives (FP), while recall is the division of true positives over the true positives and false negatives (FN).

\begin{equation}
   \small{ Precision = \frac {TP}{TP + FP}}
\end{equation}

\begin{equation}
    \small{Recall = \frac {TP}{TP + FN}}
\end{equation}

\begin{comment}
\begin{equation}
    TPR = \frac{TP}{TP + FN}    
\end{equation}
\begin{equation}
    TNR = \frac{TN}{TN + FP}
\end{equation}
\end{comment}

\hlc[highlight]{ChYP did not report any false positives, but it did report 16 false negatives, all of them vulnerabilities appearing in natural language statements, a source of information that this tool does not utilize but MDE-ChYP does. 
An example is provided in Listing} \ref{fn_tool1}.
\hlc[highlight]{Line 3 in the listing shows that \emph{`theAlarm'} device id is used to access the alarm capability of the device. 
The alarm capability in SmartThings permission model has one attribute: the \emph{`alarm'} attribute.
Lines 9 and 10 of the listing show \emph{`theAlarm'} id subscribing to changes in the \emph{`contact'} and \emph{`motion'} attributes, which are not part of the alarm capability. 
The static analysis tool did not catch these vulnerabilities as they appeared in natural language statements.}

\hlc[highlight]{MDE-ChYP reported 7 false positives due to ambiguities in analysing the code as free-form text.
Listing} \ref{fp_tool2}
\hlc[highlight]{provides an example.
This case appears when the variables are named as attributes or commands from other capabilities.
Lines 3, 8 and 9 in the listing show the variable named \emph{`alarm'} being matched with the capability \emph{`waterSensor'} and its attribute \emph{`water'}.
This confuses MDE-ChYP and reports it as malicious.
MDE-ChYP also produced 8 false negatives probably due to implementation issues that need further investigation in future work.}

\begin{lstlisting}[style=Groovy, caption={Case 1 of Over-Privilege Found in Natural Language Statements}, label=fn_tool1]
preferences {
    section("Notify me when there is any activity on this alarm:") {
        input *"theAlarm"*, *"capability.alarm"*, multiple: false, required: true
    }
}
...
def initialize() {
    log.debug "in initialize"
    subscribe(*theAlarm*, ~"contact"~, contactTriggered)
    subscribe(*theAlarm*, ~"motion"~, motionTriggered)
}
\end{lstlisting}

\begin{lstlisting}[style=Groovy, caption={Example of a False Positive Reported by the MDE based tool}, label={fp_tool2}]
preferences {
	section("When there's water detected...") {
		input *"alarm"*, ~"capability.waterSensor"~, title: "Where?"
	}
}

def installed() {
	subscribe(*alarm*, ~"water.wet"~, waterWetHandler)
	subscribe(*alarm*, ~"water.dry"~, waterWetHandler)
}
\end{lstlisting}

\begin{comment}
The first row in table \ref{table:evaluation}
shows that MDE-ChYP achieved a 78.12\% precision in detecting case 1 of over-privilege in 25 apps from the dataset. On the other hand, ChYP achieved a 100\% precision when it detected 76 over-privileges of case 1 in 222 apps, see Table \ref{table:comparison}. 
The reason that MDE-ChYP is less accurate is due to the semantics analysis done in the approach, which is more prone to error. 
Contrarily, ChYP has bounded pre-defined lists in its TXL grammar that the tool consults in its analysis. 
Leaving the chances of error to the bugs in the implementation.
\end{comment}

\hlc[highlight]{The second aspect in this comparison is the tools' coverage in detecting the three cases of over-privilege in SmartThings apps.
Due to the nature of MDE-ChYP tool and approach, it can detect over-privileges resulted from confusions in the semantics.
The static analysis based tool can only detect over-privileges exposed from analysing the syntax.
This is why the MDE approach can detect cases 1, 2 and 3 of over-privilege in SmartThings apps, while the static analysis approach only detects case 1.}

\begin{table}[h!]
    \caption{\hlc[highlight]{Evaluation comparison between the MDE and the static analysis based tools}}
    \label{table:comparison}
    \begin{tabular}{ccc}
    \toprule
    Tool                       & Precision & Recall  \\
    \midrule
    Static Analysis Tool(ChYP) & 100\%     & 60.98\% \\
    MDE-ChYP             & 82.50\%   & 80.49\% \\
    \bottomrule
\end{tabular}
\end{table}

\subsection{Manual Validation}\label{sec:rq2}
In section \ref{sec:rq1}, \hlc[highlight]{we present the approach results when evaluated on the whole dataset. The approach reported over-privilege occurrences of all three cases as shown in Table} \ref{table:initial_running_resutls}.

\hlc[highlight]{In order to quantitatively validate our findings and to compute precision and recall, we need to compare our results with a base tool, however, and since we did not have access to any tool that provides an analysis for the three types of over-privilege, we need to either manually validate all our results or randomly select random samples from the results and compute precision and recall. Validating all the results manually is time consuming, and selecting random subset from the results does not always guarantee the selected sample has all the over-privilege cases we aim to evaluate. We have already observed that from the previous section,} \ref{sec:rq3}, \hlc[highlight]{when we compared the MDE approach with ChYP for case1 of over-privilege. For this reason, we needed to create a benchmark for the evaluation using mutation analysis.}
%After running our tool on the dataset, over-privilege occurrences were reported for all three cases (see Table \ref{table:initial_running_resutls}), however, in the absence of comparable tools to compare with and in order to compute precision and recall, a manual validation of the results is required. %were not enough to evaluate the tool's performance. 
%\hlc[highlight]{For some of the cases,} manual validation of the results is time consuming, 
%\hlc[highlight]{as can be seen in Table} \ref{table:comparison_detailed},
%so we created our own benchmark by injecting over-privilege vulnerabilities.
\subsubsection{\hlc[highlight]{Benchmark}}
\hlc[highlight]{We have selected a sample benign dataset, and we manually confirmed the dataset is benign, then in each app in the dataset, we injected one case of over-privilege. Mutants created for case 1 required actuating commands and attributes that do not belong to the requested capability.}
\begin{lstlisting}[style=Groovy, caption={App Before Injecting Over-privilege}, label= before_injection]
definition(
 name: "Big Turn OFF",
 namespace: "smartthings",
 author: "SmartThings",
description: *"Turn your lights off when the SmartApp is tapped"*,
    ...)
preferences {
	section("When I touch the app, turn off...") {
		input *"switches"*, *"capability.switch"*, multiple: true
	}}
def installed()
{   subscribe(location, changedLocationMode)
	subscribe(app, appTouch) }
...
def appTouch(evt) {
	log.debug "appTouch: $evt"
	switches?.off()}
\end{lstlisting}
Listing \ref{before_injection} provides an example of an app with no over-privileges of any case.
The app is modified in listing \ref{case_1_after_injection} to show an example of an app with over-privilege of case 1. The app used the \emph{`switch'} device to access the command \emph{`siren'}, which belongs to another capability (line 18 of listing \ref{case_1_after_injection}).

\begin{lstlisting}[style=Groovy, caption={App After Injecting Case 1 of Over-privilege}, label= case_1_after_injection]
definition(
    name: "Big Turn OFF",
    namespace: "smartthings",
    author: "SmartThings",
    description: "Turn your lights off when the SmartApp is tapped",
    ...)
preferences {
	section("When I touch the app, turn off...") {
		input *"switches"*, *"capability.switch"*, multiple: true
	}}
def installed()
{   subscribe(location, changedLocationMode)
	subscribe(app, appTouch) }
...
def appTouch(evt) {
	log.debug "appTouch: $evt"
	switches?.off()
	`switches?.siren()`}
\end{lstlisting}
To create the mutant for case 2, we randomly choose any one of the capabilities that are not mentioned in the description of the app, and we request it by the application. 

%Listing \ref{before_injection} provides an example of an app with no over-privilege of case 2. 
The app in listing \ref{before_injection} only described the use of the switch capability, which was requested by the app correctly. 
To inject case 2 of over-privilege, we find a capability from the SmartThings documentation that is not supposed to be requested by the app based on its description. 
Then, we request this capability in the app without informing the user in the description. 
Listing \ref{capabilities_list} shows some of the capabilities in the SmartThings documentation. 
The resulting over-privileged app is shown in listing \ref{case2_inject_2}, line 11 shows the app requested the capability \emph{accelerationSensor} without mentioning it in the description in line 5.
%\begin{minipage}{\linewidth}
%\end{minipage}
%\begin{minipage}{\linewidth}
\begin{lstlisting}[style=Groovy, caption={Snippet of SmartThings List of Capabilities}, label= capabilities_list]
Capabilities: accelerationSensor, alarm, audioNotification, battery, beacon, bulb, button,carbonDioxideMeasurement, colorControl, colorTemperature, ...
\end{lstlisting}
%\end{minipage}
%\begin{minipage}{\linewidth}
\begin{lstlisting}[style=Groovy, caption={App After Injecting Case 2 of Over-privilege}, label= case2_inject_2]
definition(
 name: "Big Turn OFF",
 namespace: "smartthings",
 author: "SmartThings",
 description: *"Turn your lights off when the SmartApp is tapped"*,
    ...)
preferences {
 section("When I touch the app, turn off...") {
  input "switches", "capability.switch", multiple: true}
  ~section("") {~
  ~input `"sensor"`, `"capability.accelerationSensor"`~
                            ~ , multiple: true}~
}
\end{lstlisting}
%\end{minipage}
We created 19 mutants for case 2 in 19 apps from the dataset. 
As for case 3 of over-privilege, we picked apps from the dataset that have requested at least one capability, and described the use of part of it only in the app's description. 

Listing \ref{before_injection} shows a compatible app for injecting case 3. This app describes the use of the command ``off'' from the switch capability. To inject case 3 of over-privilege, we replace the actuating of the ``off'' command to the ``on'' command that is part of the switch capability. Listing \ref{case_3_after_injection} shows the app after the injection of case 3. We created 20 mutants in 20 apps from the dataset.

%\begin{minipage}{\linewidth}
\begin{lstlisting}[style=Groovy, caption={App After Injecting Case 3 of Over-privilege}, label= case_3_after_injection]
definition(
 name: "Big Turn OFF",
 namespace: "smartthings",
 author: "SmartThings",
 description: *"Turn your lights off when the SmartApp is tapped"*,
    ...)
preferences {
	section("When I touch the app, turn off...") {
		input *"switches"*, *"capability.switch"*, multiple: true
	}}
def installed()
{   subscribe(location, changedLocationMode)
	subscribe(app, appTouch) }
...
def appTouch(evt) {
	log.debug "appTouch: $evt"
    switches?.off()
	`switches?.on()`}
\end{lstlisting}
%\end{minipage}

\hlc[highlight]{For further details, the complete benchmark could be found online, as well as the public datasets. \footnote{\href{https://github.com/SmartThingsOverprivilege/smartthings_overprivilege_dataset}{github.com/SmartThingsOverprivilege/smartthings\_overprivilege\_dataset}}}

\hlc[highlight]{To answer RQ3, we run the tool on a selected subset of the dataset and the benchmark we created. 
For case 1 of over-privilege, we are interested in detecting the use of commands and attributes by capabilities that are not in possession of them.
We randomly picked 25 over-privileged apps from the public datasets (listed in Table} \ref{table:initial_running_resutls})
\hlc[highlight]{and validated the reports of case 1 of over-privilege, a detailed analysis for case1 was discussed in the previous section.
For case 2, we evaluate the tool in reporting the use of extra unneeded capabilities and in the use of the expected capabilities. We achieve this by evaluating the results obtained from running the tool on malicious and benign datasets.
For case 3, we are also interested in evaluating the tool on malicious and benign datasets. The use of a command or attribute from the owning capability should only be reported in case it is unnecessary.
%The process of the evaluation experiments for "case 2 malicious" and "cases 3 malicious and benign" turned out %to be time consuming and sometimes there are not enough true positives to report. This led us to create our own %benchmark for the evaluation. 
Finally, to evaluate the accuracy of the tool, we manually check the results obtained from the automated detection and calculate the precision and the recall.}

Table \ref{table:evaluation} presents the \hlc[highlight]{statistical validation} results for our analysis. 
Our tool could detect over-privilege vulnerabilities in malicious apps with high precision and recall. 
%While analysis on benign apps for cases 2 and 3 gave 100\% precision and 39\%-82\% recall. 
The lowest recall percentage appears in analyzing benign apps for case 3, where high numbers of false negatives are being reported. %Future work can be done on decreasing the false positives.
%\hlc[highlight]{Reasons and possible solutions are discussed in section} \ref{sec:discussion}.

% \usepackage{multirow}
\begin{table*}[t!]
\centering
  \caption{Evaluation of MDE-ChYP  on the mutation dataset}
  \label{table:evaluation}
\begin{tabular}{cccccccc}
\hline
Dataset            & \begin{tabular}[c]{@{}c@{}}Total Apps \\ Evaluated\end{tabular} & \begin{tabular}[c]{@{}c@{}}Case Occurrences\\ in dataset\end{tabular} & TP & FP & FN & Precision                & Recall                   \\ \hline
Malicious - Case 1 & 25                                                              & 41                                                                    & 33 & 7  & 8  & 82.50\%                  & 80.49\%                  \\ \hline
Malicious - Case 2 & 19                                                              & 19                                                                    & 19 & 8  & 0  & \multirow{2}{*}{92.66\%} & \multirow{2}{*}{74.81\%} \\
Benign - Case 2    & 60                                                              & 0                                                                     & 82 & 0  & 34 &                          &                          \\ \hline
Malicious - Case 3 & 20                                                              & 20                                                                    & 19 & 8  & 1  & \multirow{2}{*}{82.98\%} & \multirow{2}{*}{54.93\%} \\
Benign - Case 3    & 20                                                              & 0                                                                     & 20 & 0  & 31 &                          &                          \\ \hline
\end{tabular}
\end{table*}

\hlc[highlight]{We investigated the apps evaluated %in section} \ref{sec:rq2}
for the reasons of false positives and false negatives and pinpointed some reasons of error and areas for improvements. Fist, our NLP approach could be improved by exploring better tools and techniques in order to better understand the context and in turn produce a better match. NLP pre-processing techniques involving punctuation is one area that could immediately improve the results.
Another possible improvement is expanding the vocabulary in the tool produced using grammar inference. 

Nevertheless, even if we achieve all the desired improvements in the tool, the description could be written poorly and ambiguously that even the best tools could not comprehend the necessary privileges. Developers are strongly advised to write clear and comprehensive descriptions. This is one way we can reduce over-privileges in apps. 

Another way of mitigation is designing the access control component in the programming framework to be the most fine-grained, and ask the user for approval as straightforward as it should be. Of course, an obvious mitigation is to resolve design flaws that result in over-privileges, like case 1 in SmartThings.}

\section{Related Work}
\begin{comment}
Existing work on the detection of over-privilege in SmartThings include two methods:
\begin{itemize}
    \item Detecting over-privilege and patching the application \cite{TianZLWUGT17}
    \item Executing the application and monitoring its behavior, then comparing it to the extracted behavior from the source code or description in the UI (if the source code is not available) \cite{zhang2018homonit} 
\end{itemize}
Our solution targets the security auditors of the platform, or the third-party app developers. Where access to the source code is available and analysis can be done statically instead of both statically and dynamically as in \cite{zhang2018homonit}. Once over-privilege is detected in our solution, the developers can modify the app's source code instead of patching the app by our tool, as in \cite{TianZLWUGT17}.
\end{comment}

%\subsubsection{Security Analysis of Emerging Smart Home Applications}
Fernandes \emph{et al.} \cite{FernandesJP16} performed the first empirical security analysis of the SmartThings platform, and provided lessons for the design of smart home programming frameworks. They presented in detail the different types of vulnerabilities found in the SmartThings platform, and conducted proof-of-concept attacks by exploiting the design flaws in the platform. This was a starting point for us to understand how over-privilege results in a SmartApp.

Fernandes \emph{et al.} developed a static analysis tool that detects over-privilege in SmartThings, in addition to run-time analysis and manual analysis where the tool fails to complete the analysis. The tool is developed to detect cases 1 \& 3 of over-privilege in SmartThings. They used the tool to better understand the extent to which SmartApps are over-privileged. However, they did not evaluate the effectiveness of the developed tool, nor did they provide an easy way to evaluate it by researchers.

%\subsubsection{HoMonit}
HoMonit, a system designed and developed by Zhang \emph{et al.} \cite{zhang2018homonit}, detects two types of vulnerabilities in the SmartThings platform: over-privilege in SmartApps and event eavesdropping and spoofing. HoMonit first extracts the expected behavior of a SmartApp either by static analysis of open-source SmartApps or by NLP techniques on closed-source SmartApps. Then, it performs side-channel analysis to monitor the size and interval of the encrypted packets. Changes in the sniffed packets between benign and malicious apps indicate a change in the DFA state. From this change, it can be inferred, with high probability, that the app is not behaving as expected.

Zhang \emph{et al.} evaluated HoMonit and received a 0.98 rate of correctly labeling misbehaving SmartApps based on over-privilege access. Although, over-privilege detection in HoMonit only targets case 3 (caused by coarse-grained capabilities) and requires executing the apps in the benign and malicious states to compare between them.

%\subsubsection{SmartAuth}
Tian \emph{et al.} \cite{TianZLWUGT17} presented a technique named \emph{SmartAuth} that provides a new solution to the over-privilege problem in IoT. 
SmartAuth proposes a user-centric authorization through the generation of new authorization user interfaces based on what the app actually performs. 
SmartAuth uses NLP and program analysis to analyze an app's description, code and annotations to detect over-privilege in the app. 

To conduct the analysis, SmartAuth first analyzes the description using NLP and then analyzes the app code using NLP and program analysis. The extracted privileges from the description are compared with the privileges extracted from the app code to decide which privileges are not properly requested by the app.
SmartAuth detects cases 1 \& 3 of over-privilege, then patches the malicious app at run-time. The unavailability of SmartAuth tool publicly and the dataset used makes it hard to compare the efficiency of SmartAuth with our tool.

Our approach targets the security auditors of the platform, or the third-party app developers. Where access to the source code is available and analysis can be done statically instead of both statically and dynamically as in \cite{zhang2018homonit}. Once over-privilege is detected in our solution, the developers can modify the app's source code instead of patching the app by our tool, as in \cite{TianZLWUGT17}.

%\subsubsection{IoTCom}
IoTCom is an approach and a tool developed to detect threats at the interaction level between IoT apps \cite{alhanahnah2019advanced}. 
In this study, they present a smart-home automation model that defines an app as a set of one or more rules. Each rule can have a set of triggers, conditions and actions.
This model is used to extract the apps' behaviours using static analysis, after that they perform formal analysis to detect threats in IoT apps interactions. The static analysis performed is path-sensitive which accounts for conditions in the extracted rules.

SmartHomeML is a domain specific modelling language (DSML) used to generate applications for smart-home platforms,
specifically for Alexa and SmartThings \cite{einarsson2017smarthomeml}.
SmartHomeML adopts Model Driven Engineering (MDE) techniques for the rapid generation of the apps. In this study, they designed a meta-model that abstracts the structure of applications in different platforms, and used it to generate those applications automatically.

Our MDE approach extract permission models from the applications, our experiment finds MDE to be a great option for decreasing the complexity of extracting the expected permissions and behaviour of an application and that by analysing various aspects of SmartApps, this includes code, free-form text and user preferences. Unlike the above techniques, Our approach  can detect the three cases of SmarApps over-privilege.% in SmartApps.

%Table \ref{table:related_work} summarizes the comparison between our work and related ones.
%\begin{comment}
\begin{table*}[t!]
\centering
  \caption{Comparison with Related Work}
  \label{table:related_work}
  \resizebox{0.9\textwidth}{!}{%
  \begin{tabular}{lll}
    \toprule
    Paper                                                                                            & What it does                                                                            & Techniques Employed                                                                                                                                                   \\
    \midrule
    \begin{tabular}[l]{@{}l@{}}Security Analysis of Emerging \\ Smart Home Applications. \citet{FernandesJP16}\end{tabular} & \begin{tabular}[l]{@{}l@{}}Over-privilege detection:\\ Cases 1 \& 3\end{tabular}        & \begin{tabular}[l]{@{}l@{}}Static analysis\\ Runtime analysis\\ Manual analysis\end{tabular}                                                                          \\ \hline
    HoMonit. \citet{zhang2018homonit}                                                              & \begin{tabular}[l]{@{}l@{}}Over-privilege detection: \\ Case 3\end{tabular}             & \begin{tabular}[l]{@{}l@{}}Static analysis and NLP\\ to extract DFA of\\ expected behaviour.\\ Side channel analysis\\ to extract DFA during\\ runtime\end{tabular} \\ \hline
    SmartAuth. \citet{TianZLWUGT17}                                                                                        & \begin{tabular}[l]{@{}l@{}}Over-privilege detection:\\ Cases 1 \& 3\end{tabular}        & \begin{tabular}[c]{@{}l@{}}NLP\\ Program analysis\end{tabular}                                                                                                                                                                       \\ \hline
    IoTCom. \citet{alhanahnah2019advanced}                                                                                           & \begin{tabular}[c]{@{}l@{}}Detects threats in IoT at the\\ interaction level between apps\end{tabular}                                   & \begin{tabular}[c]{@{}l@{}}Static analysis, path-sensitive\\ Formal analysis\end{tabular}                            \\ \hline
    SmartHomeML. \citet{einarsson2017smarthomeml}                                                                                      & \begin{tabular}[l]{@{}l@{}}Generate smart-home applications\\ using a DSML\end{tabular} & \begin{tabular}[l]{@{}l@{}}MDE for fast generation\\ of apps\end{tabular}                                                                                             \\ \hline
    MDE-ChYP (our tool)\footnote{https://cresset.scs.ryerson.ca/ChYP}                                                                                             & \begin{tabular}[l]{@{}l@{}}Over-privilege detection:\\ Cases 1,2 \& 3\end{tabular}      & \begin{tabular}[l]{@{}l@{}}Static analysis to detect\\ over-privilege.\\ MDE for extracting expected\\ and actual behaviors of apps\end{tabular}                    \\ 
    \bottomrule
  \end{tabular}}
\end{table*}
%\end{comment}
%\vspace{-.2}

\hlc[highlight]{Table 6 summarizes how our tool compares to related work, including our previous tool.}

\section{Conclusions \& Future Work}
%\vspace{-.2}
%This paper presents  the design and implementation of a static analysis approach for the detection of over-privilege in SmartThings. This approach is beneficial when analyzing software by syntax and performing pattern matching. Experimental results show that this approach and tool reach 100\% precision. Assuming we have access to the source code of the apps and the documented permission model of the smart-home platform.
%we apply model driven engineering (MDE), a methodology used to abstract the system and extract models that represent it at a high level, through a process called meta-modeling \cite{schmidt2006model}. Those models and constraints allow for the process of model checking. We chose Prolog for the model checking, through applying Prolog rules and queries against Prolog facts to conduct the detection of over-privilege.
%We demonstrate how the MDE approach is capable of combining multiple sources of information for better understanding of permissions granted to the software.
%This paper presents the design and development of a MDE approach and a tool  \footnote{ https://141.117.231.79/PEMD} for the detection of over-privilege in SmartThings. This approach takes into consideration the semantics of the free-form text in the software, which allows for better understanding of the intended permission model in the software, which in turn gives better coverage of over-privilege detection. Experimental results give 78.12\% precision and 92.59\% recall for case 1,
%88.24\% precision and 78.95 \% recall for case 2, 70.37\% precision and 95\% recall for case 3.
This paper presents the design and development of an MDE approach and a tool  \footnote{https://cresset.scs.ryerson.ca/ChYP} for the detection of over-privilege in SmartThings. \hlc[highlight]{MDE is used to abstract the system and extract models that represent it at a high level}. Those models and constraints allow for the process of model checking. We chose Prolog for the model checking, through applying Prolog rules and queries against Prolog facts to conduct the detection of over-privilege.
We demonstrate how our approach is capable of combining multiple sources of information for better understanding of permissions granted to the software. This approach takes into consideration the semantics of the free-form text in the software, which allows for better understanding of the intended permission model in the software, which in turn gives better coverage of over-privilege detection. Experimental results give 78.12\% precision and 92.59\% recall for case 1, 88.24\% precision and 78.95 \% recall for case 2, 70.37\% precision and 95\% recall for case 3. Our plans for future work include working on decreasing the false positives in the detection of over-privilege. We intend to do this by working on the following aspects: investigating how to fix limitations introduced from the chosen programming paradigm of TXL's and improving the natural language understanding in our approach. %satisfies the implementation of the tool,

%\section{Future Work}\label{futurework}
%*** End of chapter: conclusions

%\appendix
%\appendixpage
%\input{Appendices/Appendix1}
 % \section{Prolog Rules for Case 1 of Over-Privilege}\label{appendix1}
 % %\addcontentsline{loaf}{section}{\protect\numberline{A} Prolog Rules for Case 1}
  %  \lstinputlisting[style=Prolog-pygsty, caption=Prolog Rules for Case 1]{"./Code/PrologCase1.pl"}

%%
%\input{Appendices/Appendix1}
  \section{Appendix A: Prolog Rules for Case 1 of Over-Privilege}\label{appendix1}
  \addcontentsline{loaf}{section}{\protect\numberline{A} Prolog Rules for Case 1}
\begin{lstlisting}[style=Prolog-pygsty, caption=Prolog Rules for Case 1]
    /* Prolog rule of case 1 */
    overprivilegedCase1(case1,AppName,RuleId,Id,Capability,Resource):-
    triggerComposition(code,AppName,RuleId,Id,Capability,Resource,_),
    notAttributeCommandOfCapability(Capability,Resource),
    attributeCommandOfCapability(Capability2,Resource),
    not(Capability2=Capability),
    not(Capability=na),
    not(Resource=na);

    triggerComposition(code,AppName,RuleId,Id,Capability,_,Resource),
    not(valueOfAttributeOfCapability(Capability,Resource)),
    valueOfAttributeOfCapability(Capability2,Resource),
    not(Capability2=Capability),
    not(Capability=na),
    not(Resource=na);
    
    actionComposition(code,AppName,RuleId,Id,Capability,Resource,_),
    notAttributeCommandOfCapability(Capability,Resource),
    attributeCommandOfCapability(Capability2,Resource),
    not(Capability2=Capability),
    not(Capability=na),
    not(Resource=na);

    actionComposition(code,AppName,RuleId,Id,Capability,_,Resource),
    not(valueOfAttributeOfCapability(Capability,Resource)),
    valueOfAttributeOfCapability(Capability2,Resource),
    not(Capability2=Capability),
    not(Capability=na),
    not(Resource=na);

    conditionComposition(code,AppName,RuleId,Id,Capability,Resource,_),
    notAttributeCommandOfCapability(Capability,Resource),
    attributeCommandOfCapability(Capability2,Resource),
    not(Capability2=Capability),
    not(Capability=na),
    not(Resource=na);

    conditionComposition(code,AppName,RuleId,Id,Capability,_,Resource),
    not(valueOfAttributeOfCapability(Capability,Resource)),
    valueOfAttributeOfCapability(Capability2,Resource),
    not(Capability2=Capability),
    not(Capability=na),
    not(Resource=na).
    /* End of rule */
    
    /* Helper Sub-Rules */
    attributeCommandOfCapability(Capability,Resource):-
    capability(Capability),
    attributeCommandOf(Capability,Resource).

    notAttributeCommandOfCapability(Capability,Resource):-
    capability(Capability),
    not(attributeCommandOf(Capability,Resource)).

    valueOfAttributeOfCapability(Capability,Resource):-
    capability(Capability),
    attributeCommandOf(Capability,Attribute),
    valueOf(Attribute,Resource).
\end{lstlisting}

%\begin{acks}
% This work is supported in part by the Natural Sciences and %Engineering Research Council of Canada (NSERC), and Ryerson %University Faculty of Science Dean's Research Fund.
%\end{acks}
\bibliographystyle{plainnat}

\bibliography{references}
\end{document}